\documentclass[12pt]{article}

\usepackage{amsfonts,amssymb,cite,bm,amsmath}
\usepackage[dvips]{graphicx}
\usepackage[dvips]{psfrag}

\newcommand {\beq}{\begin{equation}}
\newcommand {\eeq}{\end{equation}}
\newcommand {\beqa}{\begin{eqnarray}}
\newcommand {\eeqa}{\end{eqnarray}}
\newcommand {\n}{\nonumber \\}

\renewcommand{\theequation}{\thesection.\arabic{equation}}

\newcommand{\Ad}{{\rm Ad}}
\newcommand{\Tr}{{\rm Tr}}
\newcommand{\tr}{{\rm tr}}
\newcommand{\cL}{{\cal L}}

\newcommand{\cC}{{\cal C}}

\newcommand{\bi}{\bm i}
\newcommand{\bj}{\bm j}
\newcommand{\bk}{\bm k}

\allowdisplaybreaks

\begin{document}
\setlength{\oddsidemargin}{0cm}
\setlength{\baselineskip}{6mm}

\begin{titlepage}
\renewcommand{\thefootnote}{\fnsymbol{footnote}}
\begin{normalsize}
\begin{flushright}
\begin{tabular}{l}
OU-HET 599\\
February 2008
\end{tabular}
\end{flushright}
  \end{normalsize}

~~\\

\vspace*{0cm}
    \begin{Large}
       \begin{center}
         {Fiber Bundles and Matrix Models}
       \end{center}
    \end{Large}
\vspace{1cm}

\begin{center}
Takaaki I{\sc shii}\footnote
            {
e-mail address : 
ishii@het.phys.sci.osaka-u.ac.jp},
           Goro I{\sc shiki}\footnote
            {
e-mail address : 
ishiki@het.phys.sci.osaka-u.ac.jp},
Shinji S{\sc himasaki}\footnote
            {
e-mail address : 
shinji@het.phys.sci.osaka-u.ac.jp}
    {\sc and}
           Asato T{\sc suchiya}\footnote
           {
e-mail address : tsuchiya@het.phys.sci.osaka-u.ac.jp, 
                 address after April 2008 : Department of Physics, Shizuoka University, 836 Ohya, Suruga-ku,
                 Shizuoka 422-8529, Japan}\\
      \vspace{1cm}
                    
         {\it Department of Physics, Graduate School of Science}\\
         {\it Osaka University, Toyonaka, Osaka 560-0043, Japan}
      
\end{center}

\vspace{1cm}

\begin{abstract}
\noindent
We investigate relationship between a gauge theory on a
principal bundle and that on its base space. In the case where the principal bundle is itself
a group manifold, we also study relations of those gauge theories with a matrix model
obtained by dimensionally reducing them to zero dimensions.
First, we develop the dimensional reduction of Yang-Mills (YM) on the total space to YM-higgs 
on the base space for a general principal bundle. 
Second, we show a relationship that YM on an $SU(2)$ bundle 
is equivalent to the theory around a certain background of 
YM-higgs on its base space.
This is  an extension of our previous work \cite{IIST}, in which 
the same relationship concerning a $U(1)$ bundle is shown.
We apply these results to the case of $SU(n+1)$ as the total space.  
By dimensionally reducing YM on $SU(n+1)$, we obtain YM-higgs 
on $ SU(n+1)/SU(n)\simeq S^{2n+1}$ 
and on $SU(n+1)/(SU(n)\times U(1))\simeq CP^n $ and a matrix model.
We show that the theory around each monopole vacuum of YM-higgs on $CP^n$
is equivalent to the theory around a certain vacuum of the matrix model in the commutative limit.
By combining this with the relationship concerning a $U(1)$ bundle, we realize YM-higgs on 
$SU(n+1)/SU(n)\simeq S^{2n+1}$ in the matrix model. 
We see that the relationship concerning a $U(1)$ bundle can be interpreted as Buscher's T-duality.
\end{abstract}
\vfill
\end{titlepage}
\vfil\eject

\setcounter{footnote}{0}

\tableofcontents

\section{Introduction and conclusion}
\setcounter{equation}{0}
\renewcommand{\thefootnote}{\arabic{footnote}} 
Emergence of space-time is one of the key concepts in matrix models 
as nonperturbative definition of superstring \cite{BFSS,IKKT,DVV}. 
This phenomenon was first observed in the relationship between a gauge theory
and a matrix model.
This is the so-called large N reduction \cite{EK}. 
It states that a large $N$ planar gauge theory
is equivalent to the matrix model
that is its dimensional reduction to zero dimensions 
unless the $U(1)^D$ symmetry
is broken, where $D$ denotes the dimensionality of the original gauge theory.
However, the $U(1)^D$ symmetry is in general 
spontaneously broken for $D>2$.
There are two improved versions of the large $N$ reduced model
that preserve the $U(1)^D$ symmetry.
One is the quenched reduced model \cite{Bhanot:1982sh,Parisi:1982gp,
Gross:1982at,Das:1982ux}. 
The other is the twisted reduced 
model \cite{GonzalezArroyo:1982hz}, 
which was later rediscovered in the context of 
the noncommutative field theories \cite{Aoki:1999vr}.  
The T-duality for D-brane effective theories
\cite{Taylor:1996ik}, which we call the matrix T-duality in this
paper, share the same idea with the large N reduced model.
The statement of the matrix T-duality is that $U(N)$ Yang-Mills (YM) on $R^p\times S^1$ is
equivalent to $U(N\times\infty)$ YM-higgs on $R^p$ which is a dimensional reduction of $U(N\times\infty)$ YM
on $R^p\times S^1$ if a periodicity (orbifolding) condition is imposed.
Also, deconstruction \cite{ArkaniHamed:2001nc} and supersymmetric lattice gauge theories inspired by it 
\cite{Kaplan:2002wv}
are analogs of the matrix T-duality.
The above developments are all concerning gauge theories on flat space-time.
It is important 
to understand how gauge theories on curved space-time are realized 
in matrix models or gauge theories in lower dimensions, because it would lead us 
to gain some insights into how curved space-time is realized in matrix models as nonperturbative 
definition of superstring. Note that an interesting
approach to the description of curved spacetime by matrices was proposed
in \cite{Hanada:2005vr}.

In \cite{ISTT}, Takayama and three of the present authors found
relationships among the $SU(2|4)$ symmetric theories.
Here the $SU(2|4)$ symmetric theories 
include ${\cal N}=4$ super Yang Mills (SYM) on $R\times S^3/Z_k$, 
2+1 SYM on $R\times S^2$ \cite{Maldacena:2002rb}
and the plane wave matrix model (PWMM) \cite{Berenstein:2002jq}. 
These theories are related by dimensional reductions and possess common features: mass gap,
discrete spectrum and many discrete vacua. From the gravity duals of those vacua proposed in \cite{Lin:2005nh},
the following relations between these theories are suggested: 
the theory around each vacuum
of 2+1 SYM on $R\times S^2$ is equivalent to the theory around a certain
vacuum of PWMM, and the theory around each vacuum
of ${\cal N}=4$ SYM on $R\times S^3/Z_k$ is equivalent to the theory around
a certain vacuum of 2+1 SYM on $R\times S^2$ with the periodicity imposed.
Combining these two equivalences, we can say that the theory around each vacuum
of ${\cal N}=4$ SYM on $R\times S^3/Z_k$ is realized in PWMM.
In \cite{ISTT}, these equivalences were shown directly on the gauge theory side.
The results in \cite{ISTT} not only serve as a nontrivial check of the gauge/gravity 
correspondence for the $SU(2|4)$ theories, but they are also interesting
in the following aspects. Much work has been already done on the 
realization of the gauge theories on the fuzzy sphere 
\cite{Madore:1991bw,Grosse:1992bm,Grosse:1995ar,CarowWatamura:1998jn} 
by matrix models
\cite{Iso:2001mg} and on the monopoles on
the fuzzy sphere 
\cite{Grosse:1995jt,Baez:1998he,Landi,CarowWatamura:2004ct,Aoki:2003ye}.
Note that the realization of the fuzzy sphere by matrix models can be viewed as an extension of the twisted reduced model 
to curved space. Here in the relation between 2+1 SYM on $R\times S^2$ and PWMM, it was manifestly shown that
the continuum limit of concentric 
fuzzy spheres correspond to multi monopoles.
The relation between ${\cal N}=4$ SYM on $R\times S^3/Z_k$ and 2+1 SYM on $R\times S^2$ 
can be regarded as an extension of 
the matrix T-duality to that on a nontrivial $U(1)$ bundle, $S^3/Z_k$, whose base space is $S^2$.
Furthermore, in \cite{IIST}, we generalized the matrix T-duality to that on an arbitrary $U(1)$ bundle.
As an application of these results, in \cite{Ishii:2007sy}, Ohta and the present authors investigated
relationships among Chern-Simons theory on a $U(1)$ bundle over
a Riemann surface, BF theory with a mass term on the Riemann surface, which is equivalent to
two-dimensional Yang-Mills on the Riemann surface,  and a matrix model.
It was discussed that the former two (topological) field theories associated with topological strings 
can be realized in the matrix model.
The results in \cite{ISTT} also suggests an interesting possibility of a nonperturbative formulation 
of ${\cal N}=4$ SYM on $R\times S^3$ by PWMM, which would lead to a nonperturbative test of the AdS/CFT correspondence.

This paper is aimed at further investigation of the above developments concerning the large $N$
reduction and the matrix T-duality on curved space.
First, we develop a dimensional reduction of YM on the total space to YM-higgs 
on the base space for a general principal bundle. This also enables us to 
dimensionally reduce YM on a group manifold to a matrix model.
Second, as an extension of the work \cite{IIST}, in the case in which the fiber is $SU(2)$, 
we show that YM on the total
space is equivalent to a certain vacuum\footnote{Throughout
this paper, we consider gauge theories on manifolds with the Euclidean signature. 
Here `vacuum' represents a configuration that gives the global minimum of the classical action.} 
of YM-higgs on the base space with the periodicity imposed.
This enables us to realize YM on an $SU(2)^k\times U(1)^l$ bundle in YM-higgs on its base space.
We apply the above results to the case of $SU(n+1)$ as the total space.  
$SU(n+1)$ is viewed as $SU(n)$ bundle over $SU(n+1)/SU(n)\simeq S^{2n+1}$ 
or $SU(n)\times U(1)$ bundle over $SU(n+1)/(SU(n)\times U(1))\simeq CP^n$, and 
$SU(n+1)/SU(n)\simeq S^{2n+1}$ is viewed as $U(1)$ bundle over $CP^n$.
By the dimensional reduction, we obtain YM-higgs on $S^{2n+1}$ and $CP^n$ and a matrix model.
We find the commutative (continuum) limit of gauge theory on fuzzy $CP^n$ 
\cite{CarowWatamura:2004ct,Alexanian:2001qj,Balachandran:2001dd,Kitazawa:2002xj,Grosse:2004wm,Dolan:2006tx}
realized in the matrix model coincides with
YM-higgs on $CP^n$. Namely, we show that the theory around each monopole vacuum of YM-higgs on $CP^n$
is equivalent to the theory around a certain vacuum of the matrix model.
By combing this with the extended matrix T-duality, we realize YM-higgs on 
$SU(n+1)/SU(n)\simeq S^{2n+1}$ in the matrix model. 
We also show that the extended matrix T-duality of the $U(1)$ case developed in \cite{IIST}
can be interpreted as Buscher's T-duality \cite{Buscher:1987sk}.

In the remainder of this section, we describe the organization of the present paper, providing 
our results in detail, and finally describe some outlook.
From the same reasoning as the case of the $SU(2|4)$ symmetric theories,  
the following relationships among YM on $S^3$, YM-higgs on $S^2$ and a matrix model hold.
These theories are related to each other by dimensional reductions. The theory around each vacuum of 
YM-higgs on $S^2$ is equivalent to the theory around a certain vacuum of the matrix model.
YM on $S^3$ is equivalent to the theory around a certain vacuum of YM-higgs on $S^2$ with the
periodicity imposed. Eventually, YM on $S^3$ is realized in the matrix model.
It can be said that our results in this paper are extension of these relationships. In section 2, we show these
relationships in order to illustrate our basic ideas. 

In section 3, we develop a dimensional reduction on a general principal fiber bundle.
We start with YM on the total space, dimensionally reduce the fiber directions 
and obtain a YM-higgs on the base space. 

In section 4, we examine a relationship between
YM on the total space and YM-higgs on the base space obtained in section 3.
In section 4.1, we first examine the transformations of the fields from a local patch to another local patch
in YM-higgs on the base space. In section 4.2, using the observation in section 4.1, we show that
when the fiber is $U(1)$ or $SU(2)$, YM on the total space is equivalent to the theory around
a certain vacuum of YM-higgs on the base space with the periodicity imposed. This vacuum is given by
multimonopole configuration on the base space. We already found the $U(1)$ case of
this equivalence in \cite{IIST}. In the $SU(2)$ case, we also use the result in section 2 
that YM on $S^3$ is realized in
the matrix model. As a generalization, we
realize YM on $SU(2)^k\times U(1)^l$ bundle in YM-higgs on its base space.
In section 4.3, as an example, we consider $S^7$ which is an $SU(2)$ bundle over $S^4$.
In Fig. 1, we summarize our results in sections 3 and 4.

\begin{figure}[tbp]
\begin{center}
\includegraphics[height=3.5cm, keepaspectratio, clip]{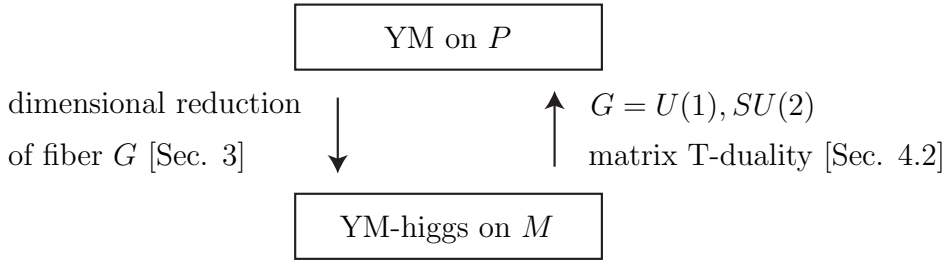}
\end{center}
\caption{Matrix T-duality for $G=U(1)$, $SU(2)$}
\end{figure}

In section 5, we examine a series of $SU(n+1)$ symmetric theories.
Fig. 2 summarizes our findings in section 5 and their relation to other sections.
The case of $n=1$ is nothing but the example discussed in section 2. In this case,
YM on $SU(2)$ is the same as YM-higgs on $S^3$ because $SU(2)\simeq S^3$.
In section 5.1, as a special case of section 3, 
we consider a dimensional reduction of YM on a group manifold $\tilde{G}$ to a coset space $\tilde{G}/H$ where
$H$ is a subgroup of $\tilde{G}$. 
Namely, we view $\tilde{G}$ as an $H$ bundle over $\tilde{G}/H$. By dimensionally reducing the Killing
vectors on $\tilde{G}$ to those on $\tilde{G}/H$, we obtain a theory on $\tilde{G}/H$ expressed in terms of
the Killing vectors. Then, we show that this theory on $\tilde{G}/H$ is rewritten into YM-higgs on 
$\tilde{G}/H$ obtained in section 3. In section 5.2, we apply the results in section 5.1 to the case of 
$\tilde{G}=SU(n+1)$ and obtain a series of theories in Fig. 2 which possess $SU(n+1)$ symmetry.
If we take $SU(n)$ as $H$, we obtain YM-higgs on $S^{2n+1}$. Note that the isometry of this $S^{2n+1}$ is not $SO(2n+2)$ but $SU(n+1)$. For $n\geq 2$, it is different from the ordinary $S^{2n+1}$ but homeomorphic to the ordinary one, and
is called a squashed $S^{2n+1}$.
If we take $SU(n)\times U(1)$ as $H$, we obtain YM-higgs on $CP^n$. Finally if we take $SU(n+1)$ itself as $H$,
we obtain a matrix model whose action is shown in Fig. 2, where $f_{ABC}$ is the structure constant of 
the $SU(n+1)$ Lie algebra. As indicated in Fig. 2, these dimensional reductions can also be performed
step by step: we obtain YM-higgs on $CP^n$ from YM-higgs on $S^{2n+1}$ and the matrix model from
YM-higgs on $CP^n$. In the case of $n=2$, as an application of the result in section 4.2, we see
that YM on $SU(3)$ is equivalent to the theory around a vacuum of YM-higgs on $S^5$ 
with the periodicity imposed ((i) in Fig. 2).
Since $S^{2n+1}$ can be viewed as a $U(1)$ bundle over $CP^n$, in section 5.3, 
we show as an application
of the results in section 4.2 that the theory around each vacuum of YM-higgs on $S^{2n+1}$ 
is equivalent to the theory around a vacuum of YM-higgs on $CP^n$ with the periodicity 
imposed ((i\hspace{-.1em}i) in Fig. 2).
In section 5.3, we show that the theory around each abelian monopole
vacuum of YM-higgs on $CP^n$ is equivalent to a certain 
vacuum of the matrix model ((i\hspace{-.1em}i\hspace{-.1em}i) in Fig. 2). Combining these results, we also
show that the theory around the trivial vacuum of 
YM-higgs on $S^{2n+1}$ is realized in the matrix model ((i\hspace{-.1em}v) in Fig. 2).
YM on $SU(3)$ is realized in YM-higgs on $CP^2$ ((v) in Fig. 2).
Finally, we make a comment: it follows from the result in section 4 
that YM on $SU(n+1)$ is realized in YM-higgs on $SU(n+1)/(SU(2)^k\times U(1)^l)$.

\begin{figure}[tbp]
\begin{center}
\includegraphics[height=9cm, keepaspectratio, clip]{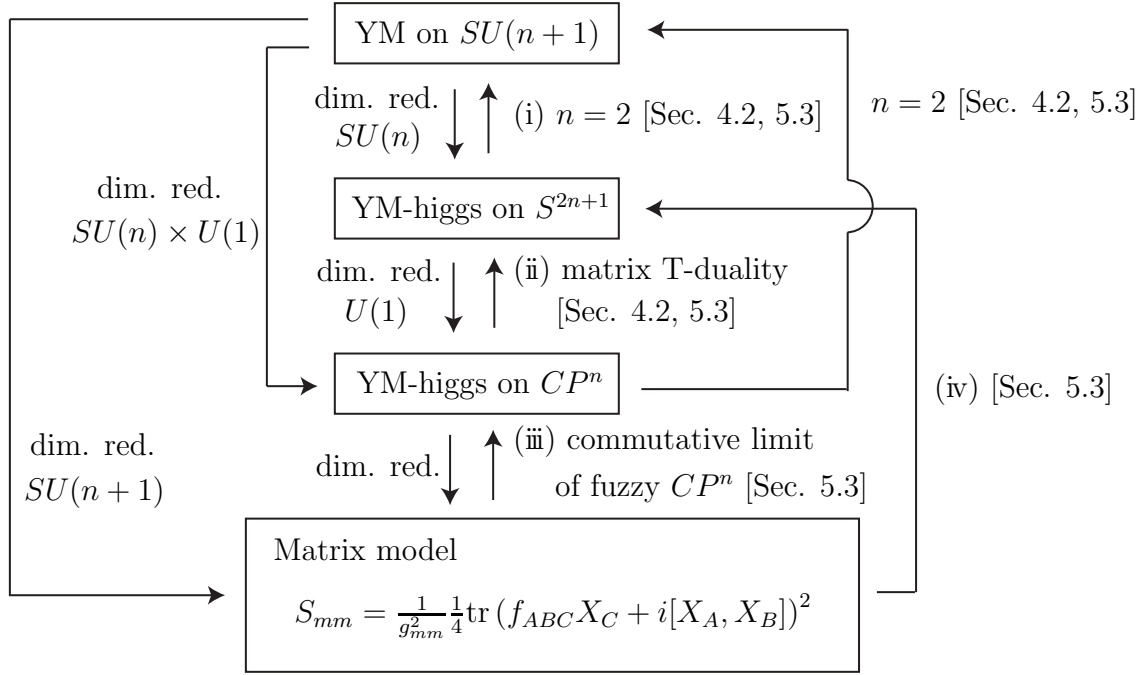}
\end{center}
\caption{A series of theories studied in section 5.}
\end{figure}


In section 6, we discuss how the extended matrix T-dulaity found in \cite{IIST} and reviewed in section 4.2
is interpreted as Buscher's T-duality.
In appendices A-D, we describe some details.

It is an open problem whether YM on $SU(n+1)$ with $n\geq 2$ is realized in the matrix model.
Presumably, we need to construct noncommutative counterparts of non-Abelian monopoles of YM-higgs on
$S^{2n+1}$ or $CP^n$ in the matrix model.
Realization of YM on $SU(n+1)$ in the matrix model should enable us to extend the matrix 
T-duality to the case of $G=SU(n+1)$. Of course, the matrix T-duality for a general $G$ should
still be investigated. It is important to see whether the matrix T-duality in the $SU(2)$ case is associated 
with the nonabelian T-duality discussed within the nonlinear sigma models \cite{Ossa:1992vc}.
It is also relevant to identify the commutative limit of the matrix model 
consisting of the square of the commutators and
the generalized Myers term with the $SU(n+1)$ structure constant which has been examined 
in \cite{Kitazawa:2002xj,Azuma:2004qe}
and find its higher-dimensional origin. 
Analysis in this paper is classical. Whether the relationships among the gauge theories
we found hold quantum mechanically is a nontrivial and important problem. 
It should be noted that in the quantum correspondence 
no orbifolding condition is needed in the matrix T-duality as far as the planar limit is concerned. 
This is nothing but the large $N$ reduction and enables us 
to make the size of matrices become finite and play a role of the ultraviolet cutoff.
In particular, we expect to give a nonperturbative definition of ${\cal N}=4$ SYM on $R\times S^3$ in 
the planar limit in terms of PWMM \cite{IIST2}.

\section{Typical relationships}
\setcounter{equation}{0}
In this section, to illustrate our ideas, we describe 
relationships among YM on $S^3$, YM-higgs on $S^2$ and a matrix model. 
These relationships are essentially the same as those among the $SU(2|4)$ symmetric
theories found in \cite{ISTT}.

We consider $S^3$ with radius $2/\mu$ and regard it as 
the $U(1)$ ($S^1$) Hopf bundle on $S^2$ with radius $1/\mu$. $S^3$ with radius $2/\mu$ is 
defined by
\begin{eqnarray}
\{(w_1,w_2)\in C^2\:|\: |w_1|^2+|w_2|^2=4/\mu^2\}.
\label{definition of S^3}
\end{eqnarray}
The Hopf map $\pi:\:S^3 \rightarrow CP^1 \;(S^2)$ is defined by
\begin{eqnarray}
(w_1,w_2) \rightarrow [(w_1,w_2)]
\equiv \{\lambda(w_1,w_2)|\lambda \in C\backslash\{0\} \}.
\end{eqnarray}
Two patches are introduced on $CP^1$: the patch I $(w_1\neq 0)$ and the 
patch II $(w_2\neq 0)$. 
On the patch I the local trivialization is given by
\begin{eqnarray}
(w_1,w_2) \rightarrow \left(\frac{w_2}{w_1},\frac{w_1}{|w_1|}\right) \in \mbox{(patch I)}\times U(1),
\label{local trivialization 1}
\end{eqnarray}
while
on the patch II the local trivialization is given by
\begin{eqnarray}
(w_1,w_2)\rightarrow \left(\frac{w_1}{w_2},\frac{w_2}{|w_2|}\right) \in \mbox{(patch II)}\times U(1).
\label{local trivialization 2}
\end{eqnarray}
The equation (\ref{definition of S^3}) is solved as 
\begin{eqnarray}
w_1=\frac{2}{\mu} \cos\frac{\theta}{2}\: e^{i\sigma_1},\;\;\;
w_2=\frac{2}{\mu} \sin\frac{\theta}{2}\: e^{i\sigma_2},
\end{eqnarray}
where $0 \leq \theta \leq \pi$ and $0 \leq \sigma_1,\:\sigma_2 < 2\pi$.
We put
\begin{eqnarray}
\varphi=\sigma_1-\sigma_2,\;\;\; \psi=\sigma_1+\sigma_2,
\end{eqnarray}
and can change the ranges of $\varphi$ and $\psi$ 
to $0\leq \varphi < 2\pi$ and $0 \leq \psi < 4\pi$.
The periodicity is expressed as 
\begin{eqnarray}
(\theta,\varphi,\psi)\sim (\theta,\varphi+2\pi,\psi+2\pi)
\sim (\theta,\varphi,\psi+4\pi).
\label{periodicity for angular variables}
\end{eqnarray}
$\mbox{}$From the local trivializations (\ref{local trivialization 1}) and 
(\ref{local trivialization 2}), one can see that 
$\theta$ and $\varphi$ are regarded as the angular coordinates of the base space
$S^2$ through the stereographic projection. 
The patch I corresponds to $0\leq \theta <\pi$, while the patch II corresponds
to $0 < \theta \leq \pi$.
The metric of $S^3$ is given as follows:
\begin{align}
ds_{S^3}^2&=|dw_1|^2+|dw_2|^2 \nonumber\\
&=\frac{1}{\mu^2}(d\theta^2+\sin^2\theta d\varphi^2+(d\psi+\cos\theta d\varphi)^2).
\label{metric of S3}
\end{align}

In the remainder of 
this section, the upper sign is taken in the patch I and the lower sign in
the patch II. From (\ref{local trivialization 1}), (\ref{local trivialization 2}) and (\ref{metric of S3}),
one sees that the fiber $S^1$ is parameterized by $y=\frac{1}{\mu}(\psi\pm \varphi)$ and its radius
is given by $2/\mu$. The connection
1-form is given by 
\begin{align}
\omega=\frac{\mu}{2}\left(dy+\frac{1}{\mu}(\cos\theta \mp 1)d\varphi\right).
\end{align} 
The connection 1-form provides the vertical-horizontal decomposition by determining
the inverse of the dreibein $E_A^M$ through $\omega(E_{\alpha}^M)=0$,
$E_3^{\mu}=0$ and $E_3^y=1$, where $A=1,2,3$, $\alpha=1,2$, $M=\theta,\varphi,y$ and $\mu=\theta,\varphi$.
The inverse of the dreibein is determined as
\begin{align}
&E_1^{\theta}=\mu, \;\;\;
E_2^{\varphi}=\frac{\mu}{\sin\theta}, \nonumber\\
&E_2^y=\mu\frac{\cos\theta\mp 1}{\sin\theta}, \;\;\;
E_3^y=1, \nonumber\\
&\mbox{others}=0.
\label{inverse of dreibein}
\end{align}
The dreibein are given by
\begin{align}
&E^1_{\theta}=e^1_{\theta}=\frac{1}{\mu}, \;\;\;
E^2_{\varphi}=e^2_{\varphi}=\frac{1}{\mu}\sin\theta, \nonumber\\
&E^3_{\varphi}=\frac{1}{\mu}(\cos\theta\mp 1), \;\;\;
E^3_y=1, \nonumber\\
&\mbox{others}=0,
\label{dreibein}
\end{align}
where $e_{\mu}^{\alpha}$ are the zweibein of $S^2$.

We start with YM on $S^3$
\begin{align}
S_{S^3}=\frac{1}{4g_{S^3}^2}\int \frac{d\Omega_3}{(\mu/2)^3} 
\mbox{tr}(F_{AB}F_{AB}).
\label{YM on S^3}
\end{align}
The vertical-horizontal decomposition tells us how to relate the gauge field on $S^3$
to the gauge field and the higgs field on $S^2$:
\begin{align}
&A_{\alpha}=a_{\alpha}, \nonumber\\
&A_3=\phi.
\label{gauge fields on S^3 and gauge fields and higgs field on S^2}
\end{align}
Or equivalently
\begin{align}
&A_{\theta}=a_{\theta} , \nonumber\\
&A_{\varphi}=a_{\varphi}+\frac{1}{\mu}(\cos\theta \mp 1)\phi, \nonumber\\
&A_{y}=\phi.
\label{gauge fields on S^3 and gauge fields and higgs field on S^2 2}
\end{align}
In (\ref{gauge fields on S^3 and gauge fields and higgs field on S^2}) and
(\ref{gauge fields on S^3 and gauge fields and higgs field on S^2 2}), in order to make a dimensional reduction,
we assume
that the both sides are independent of $y$. Then, substituting 
(\ref{gauge fields on S^3 and gauge fields and higgs field on S^2}) into 
(\ref{YM on S^3}) yields a YM-higgs on $S^2$,
\begin{align}
S_{S^2}=\frac{1}{g_{S^2}^2}\int \frac{d\Omega_2}{\mu^2} \tr \left(
\frac{1}{2}(f_{12}+\mu\phi)^2+\frac{1}{2}(D_{\alpha}\phi)^2 \right),
\label{YM-higgs on S^2}
\end{align}
where $g_{S^2}^2=\frac{\mu}{4\pi}g_{S^3}^2$. It is convenient for us to
rewrite (\ref{YM-higgs on S^2}) using the three-dimensional flat space notation.
We define a three-dimensional vector field in terms of $a_{\alpha}$ and $\phi$ \cite{Maldacena:2002rb}:
\begin{align}
\vec{X}=\phi \vec{e}_r + a_1\vec{e}_{\varphi}
-a_2\vec{e}_{\theta},
\label{X vector}
\end{align}
where 
$\vec{e}_r=(\sin\theta\cos\varphi,\sin\theta\sin\varphi,\cos\theta)$ and
$\vec{e}_{\theta}=\frac{\partial \vec{e}_r}{\partial \theta},\;\;
\vec{e}_{\varphi}=\frac{1}{\sin\theta}
\frac{\partial \vec{e}_r}{\partial \varphi}$.
We also introduce the angular momentum operator in three-dimensional flat space,
\begin{align}
\vec{L}^{(0)}=-i\vec{e}_{\phi}\partial_{\theta}
+i\frac{1}{\sin\theta}\vec{e}_{\theta}\partial_{\phi}.
\label{angular momentum operator}
\end{align}
Then, (\ref{YM-higgs on S^2}) is rewritten as  
\begin{align}
S_{S^2}=\frac{1}{g_{S^2}^2}\int \frac{d\Omega_2}{\mu^2} \frac{1}{2}
\mbox{tr}\left(\mu X_A+i\mu\epsilon_{ABC}L^{(0)}_BX_C+\frac{i}{2}\epsilon_{ABC}[X_B,X_C]\right)^2.
\label{YM-higgs on S^2 2}
\end{align}
By dropping all the derivatives, 
we dimensionally reduce (\ref{YM-higgs on S^2 2}) to zero dimensions to obtain a matrix model:
\begin{align}
S_{mm}=\frac{1}{g_{mm}^2}\frac{1}{2}\mbox{tr}\left(\mu X_A+\frac{i}{2}\epsilon_{ABC}[X_B,X_C]\right)^2,
\label{SU(2) matrix model}
\end{align}
where $g_{mm}^2=\frac{\mu^2}{4\pi}g_{S^2}^2$. 
The cross term in the above action is nothing but the Myers term \cite{Myers:1999ps}.
It was first found in \cite{Kim:2003rza} that (\ref{SU(2) matrix model}) is obtained from (\ref{YM on S^3})
through the dimensional reduction.

We can obtain (\ref{YM-higgs on S^2 2}) and (\ref{SU(2) matrix model})
directly from (\ref{YM on S^3}) in the following way. We parameterize the gauge field on $S^3$ as
$A=X_AE^A$ \cite{Lin:2005nh}, where $E^A$ is the right invariant 1-form defined in appendix A.
Then, by using the Maurer-Cartan equation (\ref{Maure-Cartan}), we evaluate the curvature 2-form as 
\begin{align}
F &=dA+iA\wedge A \nonumber\\
  &=\frac{1}{2}\epsilon_{ABC}\left(i\mu\epsilon_{CDE}{\cal L}_DX_E+\mu X_C
  +i\epsilon_{CDE}X_DX_E \right)E^A\wedge E^B,
\label{F in MC basis}
\end{align}
where ${\cal L}_A$ are the Killing vector dual to $E^A$, the explicit form of which is given 
in (\ref{Killing vector}).
Noting that ${\cal L}_A$ reduces to $L^{(0)}_A$ when $X_A$ is independent of $y$, 
one can easily see that (\ref{YM on S^3}) is dimensionally reduced to
(\ref{YM-higgs on S^2 2}). Moreover, if we assume that $X_A$ is independent of all coordinates,
we obtain the matrix model (\ref{SU(2) matrix model}) directly from (\ref{YM on S^3}).

The theories (\ref{YM-higgs on S^2}) and (\ref{SU(2) matrix model}) possess many nontrivial vacua.
Let us see how those vacua are described. 
First, the vacuum configurations 
of (\ref{YM-higgs on S^2}) with the gauge group $U(M)$ are determined by 
\begin{align}
&f_{12}+\mu\phi=0, \nonumber\\
&D_{\alpha}\phi=0.
\label{eom for YM-higgs on S^2}
\end{align}
In the gauge in which $\phi$ is diagonal, (\ref{eom for YM-higgs on S^2}) is solved as
\begin{align}
&\hat{a}_1=0, \nonumber\\
&\hat{a}_2=\frac{\cos\theta\mp 1}{\sin\theta}\hat{\phi}, \nonumber\\
&\hat{\phi}=\frac{\mu}{2}\mbox{diag}(\cdots,\underbrace{n_{s-1},\cdots,n_{s-1}}_{N_{s-1}},
\underbrace{n_{s},\cdots,n_{s}}_{N_{s}},\underbrace{n_{s+1},\cdots,n_{s+1}}_{N_{s+1}},\cdots),
\label{vacuum of YM-higgs on S^2}
\end{align}
where the gauge field takes the configurations of Dirac's monopoles, so that 
$n_s$ must be integers due to Dirac's quantization condition. Note also that $\sum_sN_s=M$.
Thus the vacua of YM-higgs on $S^2$ are classified by the monopole charges $n_s/2$ and their degeneracies $N_s$.
Next, the vacuum configurations of (\ref{SU(2) matrix model}) with the gauge group $U(\hat{M})$ 
are determined by\footnote{There is a solution to the equations of motion of the matrix model (\ref{SU(2) matrix model}), $X_A=\frac{\mu}{2}L_A$, 
which does not satisfy (\ref{eom for SU(2) matrix model}). It turns out that the theory around this solution is
unstable.}
\begin{align}
[X_A,X_B]=i\mu\epsilon_{ABC}X_C.
\label{eom for SU(2) matrix model}
\end{align}
(\ref{eom for SU(2) matrix model}) is solved as 
\begin{align}
\hat{X}_A=\mu L_A,
\label{vacuum of SU(2) matrix model}
\end{align}
where $L_A$ are the representation matrices of the $SU(2)$ generators
which are in general
reducible, and are decomposed into irreducible representations:
\begin{align}
L_A=
 \begin{pmatrix}
  \rotatebox[origin=tl]{-35}
  {$\cdots \;\;\;
  \overbrace{\rotatebox[origin=c]{35}{$L_{A}^{[j_{s-1}]}$} \;
  \cdots \;
  \rotatebox[origin=c]{35}{$L_{A}^{[j_{s-1}]}$}}^{\rotatebox{35}{$N_{s-1}$}}
  \;\;\;
  \overbrace{\rotatebox[origin=c]{35}{$L_A^{[j_{s}]}$} \;
  \cdots \;
  \rotatebox[origin=c]{35}{$L_A^{[j_{s}]}$}}^{\rotatebox{35}{$N_{s}$}}
  \;\;\;
  \overbrace{\rotatebox[origin=c]{35}{$L_{A}^{[j_{s+1}]}$} \;
  \cdots \;
  \rotatebox[origin=c]{35}{$L_{A}^{[j_{s+1}]}$}}^{\rotatebox{35}{$N_{s+1}$}}
  \;\;\;\cdots $}
 \end{pmatrix},
 \label{matrix background}
\end{align}
where $L_A^{[j]}$ are the spin $j$ representation matrices of $SU(2)$ and 
$\sum_s N_s(2j_s+1)=\hat{M}$. The vacua of the matrix model are classified by the $SU(2)$ representations $[j_s]$
and their degeneracies $N_s$. (\ref{matrix background}) represents concentric fuzzy spheres with different radii.

In the remainder of this section, we show relationships among the theories
(\ref{YM on S^3}), (\ref{YM-higgs on S^2}) and (\ref{SU(2) matrix model}).
First, we show that the theory around the vacuum (\ref{vacuum of YM-higgs on S^2}) of YM-higgs on $S^2$ is
equivalent to the theory around the vacuum (\ref{vacuum of SU(2) matrix model}) of the matrix model
if one puts $2j_s+1=N_0+n_s$ and takes the $N_0 \rightarrow \infty$ limit with $g_{mm}^2/N_0$ 
fixed to $g_{S^2}^2\mu^2/4\pi$.
We decompose the fields into the background corresponding (\ref{vacuum of YM-higgs on S^2})
and the fluctuation as $X_A^{(s,t)} \rightarrow \hat{X}_A^{(s,t)}+X_A^{(s,t)}$, where $(s,t)$
label the (off-diagonal) blocks. Then, (\ref{YM-higgs on S^2 2}) is expanded around
(\ref{vacuum of YM-higgs on S^2}) as 
\begin{align}
S_{S^2}&=\frac{1}{g_{S^2}^2}\int \frac{d\Omega_2}{\mu^2} \frac{1}{2}\sum_{s,t}
\mbox{tr}\left[\left(\mu X_A^{(s,t)}+i\mu\epsilon_{ABC}L^{(q_{st})}_BX_C^{(s,t)}
+\frac{i}{2}\epsilon_{ABC}[X_B,X_C]^{(s,t)}\right) \right.\nonumber\\
&\left. \qquad\qquad\qquad\qquad\qquad
\times\left(\mu X_A^{(t,s)}+i\mu\epsilon_{ADE}L^{(q_{ts})}_DX_E^{(t,s)}
+\frac{i}{2}\epsilon_{ADE}[X_D,X_E]^{(t,s)}\right)\right],
\label{YM-higgs on S^2 expanded around vacuum}
\end{align}
where $q_{st}=(n_s-n_t)/2$. $\vec{L}^{(q)}$ is the angular momentum operator 
in the presence of a monopole
with the magnetic charge $q$ at the origin, which takes 
the form \cite{Wu:1976ge}
\begin{align}
\vec{L}^{(q)}=\vec{L}^{(0)}-q\frac{\cos\theta\mp 1}{\sin\theta}\vec{e}_{\theta}
-q\vec{e}_r.
\label{Lq}
\end{align}
We make a harmonic expansion of (\ref{YM-higgs on S^2 expanded around vacuum})
by expanding the fluctuation in terms of the monopole vector spherical harmonics $\tilde{Y}_{JmqA}^{\rho}$ 
defined in appendix A as
\begin{align}
X_A^{(s,t)}=\sum_{\rho=0,\pm 1}\sum_{\tilde{Q}\geq |q_{st}|}
\sum_{m=-Q}^Q X_{Jm\rho}^{(s,t)}\tilde{Y}_{JmqA}^{\rho},
\label{mode expansion of X}
\end{align}
where $Q=J+\frac{(1+\rho)\rho}{2}$ and $\tilde{Q}=J-\frac{(1-\rho)\rho}{2}$.
Substituting (\ref{mode expansion of X}) into (\ref{YM-higgs on S^2 expanded around vacuum})
yields
\begin{align}
S_{S^2}&=\frac{4\pi}{g_{S^2}^2\mu^2}\mbox{tr}\left[
\frac{\mu^2}{2}\sum_{s,t}\rho^2 (J+1)^2X_{Jm\rho}^{(s,t)\dagger}X_{Jm\rho}^{(s,t)} \right.\nonumber\\
&+i\mu\sum_{s,t,u}\rho_1(J_1+1){\cal E}_{J_1m_1q_{st}\rho_1\;J_2m_2q_{tu}\rho_2\;J_3m_3q_{us}\rho_3}
X_{J_1m_1\rho_1}^{(s,t)}X_{J_2m_2\rho_2}^{(t,u)}X_{J_3m_3\rho_3}^{(u,s)} \nonumber\\
&-\frac{1}{2}\sum_{s,t,u,v}(-1)^{m-q_{su}+1}
{\cal E}_{J-mq_{us}\rho\;J_1m_1q_{st}\rho_1\;J_2m_2q_{tu}\rho_2}
{\cal E}_{Jmq_{su}\rho\;J_3m_3q_{uv}\rho_3\;J_4m_4q_{vs}\rho_4} \nonumber\\
&\left.\qquad\qquad\qquad \times
X_{J_1m_1\rho_1}^{(s,t)}X_{J_2m_2\rho_2}^{(t,u)}X_{J_3m_3\rho_3}^{(u,v)}X_{J_4m_4\rho_4}^{(v,s)}
\right],
\label{mode expansion of YM-higgs on S^2}
\end{align}
where ${\cal E}_{J_1m_1q_{st}\rho_1\;J_2m_2q_{tu}\rho_2\;J_3m_3q_{us}\rho_3}$ is defined in
(\ref{E}) and we have used (\ref{rotation}).
Similarly we decompose the matrices into the background given by (\ref{vacuum of SU(2) matrix model})
and the fluctuation as $X_i \rightarrow \hat{X}_i+X_i$ and obtain the theory around 
(\ref{vacuum of SU(2) matrix model}):
\begin{align}
S_{mm}&=\frac{1}{g_{mm}^2}\frac{1}{2}\sum_{s,t}
\mbox{tr}\left[\left(\mu X_A^{(s,t)}+i\mu\epsilon_{ABC}L_B\circ X_C^{(s,t)}
+\frac{i}{2}\epsilon_{ABC}[X_B,X_C]^{(s,t)}\right) \right.\nonumber\\
&\left. \qquad\qquad\qquad\qquad
\times\left(\mu X_A^{(t,s)}+i\mu\epsilon_{ADE}L_D\circ X_E^{(t,s)}
+\frac{i}{2}\epsilon_{ADE}[X_D,X_E]^{(t,s)}\right)\right],
\label{SU(2) matrix model expanded around vacuum}
\end{align}
where $L_A\circ$ is defined by
\begin{align}
L_A\circ X_B^{(s,t)}=L_A^{[j_s]}X_B^{(s,t)}
-X_B^{(s,t)}L_A^{[j_t]}.
\end{align}
We make a harmonic expansion for (\ref{SU(2) matrix model expanded around vacuum})
by expanding the fluctuation in terms of the fuzzy vector spherical harmonics $\hat{Y}_{Jm(j_sj_t)A}^{\rho}$ 
defined in appendix A as
\begin{align}
X_A^{(s,t)}=\sum_{\rho=0,\pm 1}\sum_{\tilde{Q}\geq |j_s-j_t|}^{j_s+j_t}
\sum_{m=-Q}^Q X_{Jm\rho}^{(s,t)}\otimes \hat{Y}_{Jm(j_sj_t)A}^{\rho}.
\label{mode expansion of X 2}
\end{align}
Since $j_s+j_t=N_0+\frac{n_s+n_t}{2}-1$, $N_0$ plays a role of the 
ultraviolet cutoff. Note also that $j_s-j_t=(n_s-n_t)/2=q_{st}$.
Substituting (\ref{mode expansion of X 2}) into (\ref{SU(2) matrix model expanded around vacuum}) yields
\begin{align}
S_{mm}&=\frac{N_0}{g_{mm}^2}\mbox{tr}\left[
\frac{\mu^2}{2}\sum_{s,t}\rho^2 (J+1)^2X_{Jm\rho}^{(s,t)\dagger}X_{Jm\rho}^{(s,t)} \right.\nonumber\\
&+i\mu\sum_{s,t,u}\rho_1(J_1+1)\hat{{\cal E}}_{J_1m_1(j_sj_t)\rho_1\;J_2m_2(j_tj_u)\rho_2\;J_3m_3(j_uj_s)\rho_3}
X_{J_1m_1\rho_1}^{(s,t)}X_{J_2m_2\rho_2}^{(t,u)}X_{J_3m_3\rho_3}^{(u,s)} \nonumber\\
&-\frac{1}{2}\sum_{s,t,u,v}(-1)^{m-q_{su}+1}
\hat{{\cal E}}_{J-m(j_uj_s)\rho\;J_1m_1(j_sj_t)\rho_1\;J_2m_2(j_tj_u)\rho_2}
\hat{{\cal E}}_{Jm(j_sj_u)\rho\;J_3m_3(j_uj_v)\rho_3\;J_4m_4(j_vj_s)\rho_4} \nonumber\\
&\left.\qquad\qquad\qquad \times
X_{J_1m_1\rho_1}^{(s,t)}X_{J_2m_2\rho_2}^{(t,u)}X_{J_3m_3\rho_3}^{(u,v)}X_{J_4m_4\rho_4}^{(v,s)}
\right],
\label{mode expansion of SU(2) matrix model}
\end{align}
where $\hat{{\cal E}}_{J_1m_1(j_sj_t)\rho_1\;J_2m_2(j_tj_u)\rho_2\;J_3m_3(j_uj_s)\rho_3}$ is defined in (\ref{E})
and we have used (\ref{rotation}).
In the $N_0\rightarrow \infty$ limit, the ultraviolet cutoff goes to infinity 
and $\hat{{\cal E}}_{J_1m_1(j_sj_t)\rho_1\;J_2m_2(j_tj_u)\rho_2\;
J_3m_3(j_uj_s)\rho_3}$ reduces to
${\cal E}_{J_1m_1q_{st}\rho_1\;J_2m_2q_{tu}\rho_2\;
J_3m_3q_{us}\rho_3}$ as shown in appendix A. Namely, this limit corresponds to
the commutative (continuum) limit of the fuzzy spheres.
Hence, in the limit in which 
$N_0\rightarrow \infty$ and $g_{mm}\rightarrow \infty$ 
such that 
$g_{mm}^2/N_0=g_{S^2}^2\mu^2/4\pi$, 
(\ref{mode expansion of SU(2) matrix model}) agrees 
with (\ref{mode expansion of YM-higgs on S^2}). We have proven our statement.

Next, we show that the theory around a certain vacuum of $U(M=N\times\infty)$ YM-higgs on $S^2$
with a periodicity condition imposed is equivalent to $U(N)$ YM on $S^3$.
This is an extension of the matrix T-duality to a nontrivial fiber bundle.
The vacuum of YM-higgs on $S^2$
we take is given by (\ref{vacuum of YM-higgs on S^2}) with $s$ running from $-\infty$ to $\infty$,
$n_s=s$ and $N_s=N$. $4\pi g_{S^2}^2/\mu $ 
is identified with the coupling constant on $S^3$, $g_{S^3}^2$. 
We decompose the fields on $S^2$ into the background and the fluctuation,
\begin{align}
&a_{\alpha} \rightarrow \hat{a}_{\alpha}+a_{\alpha}, \nonumber\\
&\phi \rightarrow \hat{\phi}+\phi,
\label{background and fluctuation}
\end{align}
and impose the periodicity (orbifolding) condition on the fluctuation,
\begin{align}
&a_{\alpha}^{(s+1,t+1)}=a_{\alpha}^{(s,t)}\equiv a_{\alpha}^{(s-t)}, \nonumber\\
&\phi^{(s+1,t+1)}=\phi^{(s,t)}\equiv \phi^{(s-t)}.
\label{periodicity condition}
\end{align}
The fluctuations are gauge-transformed from
the patch I to the patch II as \cite{IIST} 
\begin{align}
&{a'}_{\!\alpha}^{(s-t)}
=e^{-i(s-t)\phi}a_{\alpha}^{(s-t)}, \nonumber\\
&{\phi'}^{(s-t)}
=e^{-i(s-t)\phi}\phi^{(s-t)}.
\label{gauge transformation of fluctuations}
\end{align}
We make the Fourier transformation for the fluctuations
on each patch to construct the gauge field
on the total space from the fields on the base space:
\begin{align}
A_{\alpha}(\theta,\varphi,\psi)&=\sum_{w}a_{\alpha}^{(w)}(\theta,\varphi)
e^{-i\frac{\mu}{2}wy}, \nonumber\\
A_y(\theta,\varphi,\psi)&=\sum_{w}\phi^{(w)}(\theta,\varphi)
e^{-i\frac{\mu}{2}wy}.
\label{Fourier transformation}
\end{align}
We see from (\ref{gauge transformation of fluctuations}) that
the lefthand sides of (\ref{Fourier transformation}) are indeed independent of the patches.
We substitute (\ref{Fourier transformation}) into (\ref{YM-higgs on S^2 expanded around vacuum})
and divide an overall factor $\sum_s$ to extract a single
period. Then, we obtain $U(N)$ YM on $S^3$. The details of this calculation are given as a special case of
(\ref{matric T-duality for U(1)}) and (\ref{U(1) T-duality}).

Finally, combining the above two statements, we see that the theory around (\ref{vacuum of SU(2) matrix model})
of the matrix model where $s$ runs from $-\infty$ to $\infty$, $2j_s+1=N_0+s$ is equivalent to
$U(N)$ YM on $S^3$ if the $N_0\rightarrow\infty$ limit is taken with $g_{mm}^2/N_0$ fixed to $\frac{g_{S^3}^2\mu^3}{16\pi^2}$, the periodicity condition is imposed on the fluctuation on $S^2$ and
the overall factor $\Sigma_s$ is divided. In this way, $S^3$ is realized in terms of the three matrices
$X_1,\;X_2,\;X_3$.

In sections 3-5, we generalize the results in this section. We set $\mu=1$ and set all other dimensionful parameters 
to a certain constant value.

\section{Dimensional reduction on a principal bundle}
\setcounter{equation}{0}

In this section, we provide the dimensional reduction of YM 
on a principal $G$ bundle to its base space.
The case of principal $U(1)$ bundles was already given in \cite{IIST}.
Here we consider the case where $G$ is nonabelian.

First, we give a metric and a vielbein of a fiber bundle on which pure YM is defined.
We consider a principal $G$-bundle $P$ on a manifold $M$. The base space $M$
has a covering $\cal S$, and the total space has a covering 
$\{\pi^{-1}(U) | U\in {\cal S}\}$. $\pi^{-1}(U)$ is diffeomorphic to $U\times G$ by
the local trivialization. Thus it is parameterized by 
$z^M=(x^\mu,y^m)\; (\mu=1,\cdots,\dim M\; ; \; m=1,\cdots, \dim G)$,
where $x^\mu$ parameterize the local patch $U$ and $y^m$ parameterize an
element of $G$.
We assume that the connection of $P$ is expressed as 
\begin{align}
 \omega=g^{-1}(y)b(x)g(y)-i\:g^{-1}(y)dg(y). \label{connection}
\end{align}
where $g(y)\in G$, $b(x)=b_\mu^a(x)T^a dx^\mu$ and $T^a$ are the
generators of the Lie group $G$. 

The transition functions of a principal bundle act on fibers by left
multiplication. 
If there is overlap between $U$ and $U'$,
 the relation between fiber coodinates, $g(y)$ on $U$ and $g(y')$ on $U'$, 
is given by
\begin{align}
 g(y')=k(x)\: g(y) \label{transf. of g}
\end{align}
where $k(x)\in G$. 
In the overlapping region $U \cap U'$, $b(x)$ must transform as 
\begin{align}
 b'(x')=k(x)\: b(x)\: k^{-1}(x)+i\:dk(x)k^{-1}(x).\label{b' and b} 
\end{align}
Indeed, by using (\ref{b' and b}), we can show 
\begin{align}
\omega=g(y)^{-1}\:b(x)\:g(y)-i\:g(y)^{-1}dg(y)
 =g(y')^{-1}\:b'(x')\:g(y')-i\:g(y')^{-1}dg(y').
\end{align}

We assume that the total space is endowed with a metric that
has the fibered structure determined by the connection
(\ref{connection}) and the isometry.
As shown in \cite{Choquet-Bruhat}, such metric can be locally expressed as\footnote{Throughout of this paper, we
use the following normalizations for the traces: $\mbox{Tr}(T^aT^b)=\frac{1}{2}\delta_{ab}$ for the structure group
of the fiber bundle and $\mbox{tr}(T^aT^b)=\delta_{ab}$ for the gauge group.}
\begin{align}
 ds^2&=G_{MN}dz^M dz^N  \notag \\
 &=g_{\mu\nu}(x)dx^\mu dx^\nu+2{\rm Tr}\omega^2  \notag \\
 &=g_{\mu\nu}(x)dx^\mu dx^\nu+\{e_m^a(y) dy^m -b^a_\mu(x) dx^\mu\}^2.
 \label{metric}
\end{align}
Here $g_{\mu\nu}$ is a metric on the base space and 
$e_m^a(y)\; (a=\dim M+1, \cdots, \dim P)$ are the components of 
the right invariant Maurer-Cartan 1-form of $G$, 
which is defined by 
\begin{align}
dg(y)g(y)^{-1}=-i e^a_m(y)T^a dy^m. 
\end{align}
We have assumed that the coefficient of the second term in 
(\ref{metric}) is just $\delta_{ab}$ so that
the resultant dimensionally reduced theory is simple,
although it is allowed to take $y$ independent function $\xi_{ab}(x)$.
The Maurer-Cartan 1-form satisfies the Maurer-Cartan equation
\begin{align}
 de^a-\frac{1}{2}f^{abc}e^b\wedge e^c=0, \label{MC eq}
\end{align}
where $f^{abc}$ is the structure constant of the Lie algebra of $G$, and 
is regarded as the vielbein of the Cartan-Killing metric 
on $G$ defined by
\begin{align}
 h_{mn}(y)dy^m dy^n
 &\equiv -2\Tr\left(dgg^{-1}\right)^2 \notag\\
 &=e^a_m(y)e^a_n(y)dy^m dy^n.
\end{align}
Note that $e^a_m(y)$ and $b(x)$ in the metric (\ref{metric})
are defined locally on $U$ and
must be transformed from $U$ to $U'$:
the transformation of $e^a_m(y)$ is determined by (\ref{transf. of g}) 
and an equality
\begin{align}
dg(y')g(y')^{-1}=-i e^a_m(y')T^a dy'^m, \label{MC of g'}
\end{align}
while the transformation of $b(x)$ is given in (\ref{b' and b}).
By introducing a vielbein on the base space, 
$e^\alpha_\mu(x)\; (\alpha=1, \cdots, \dim M)$,
one can write a vielbein and its inverse 
on the total space as follows:
\begin{align}
 E^A_{\;\;M}(z)=
 \begin{pmatrix}
  e^{\alpha}_\mu(x) & 0 \\
  -b^{a}_\mu(x) & e^{a}_{m}(y)
 \end{pmatrix}, \quad
E^{M}_{\;\;A}(z)=
 \begin{pmatrix}
  e^{\mu}_{\alpha}(x) & 0 \\
  e^m_a(y)b^a_\alpha(x) & e^m_a(y)
 \end{pmatrix}, \label{general vielbeins}
\end{align}
where $e^{\mu}_{\alpha}$ and $e^m_a$ are the inverse of 
$e^\alpha_\mu$ and $e^a_m$, respectively,
and $b^a_\alpha(x)\equiv e^\mu_\alpha(x)b^a_\mu(x)$.
The local Lorentz frame defined by (\ref{general vielbeins}) gives
the vertical-horizontal decomposition of vectors and 1-forms on the total
space. Namely, $\alpha=1,\cdots,\dim M$ correspond to the
directions to those of the base space and $a=\dim M +1, \cdots, \dim P$
correspond of the fiber space. 
Again, we remark that
these expressions are defined locally on $U$.
From (\ref{b' and b}) and (\ref{MC of g'}), we can obtain
relationships of the vielbeins between on $U$ and on $U'$ as
\begin{align}
 &E'^\alpha=E^\alpha, \notag\\
 &E'^a=\Ad(k)^{ab}E^b. \label{transf. of E}
\end{align}
where $\Ad(k)$ is the adjoint representation of $k(x)$.
(\ref{metric of S3}) is a counterpart of (\ref{metric}), and 
(\ref{inverse of dreibein}) and (\ref{dreibein}) are a counterpart of (\ref{general vielbeins}).

We next consider a gauge theory on the total space and make a dimensional
reduction of the fiber direction to obtain a gauge theory on the base
space. We start with $U(N)$ YM on the total space:
\begin{align}
 S_P=\frac{1}{g^2_P}\int d^{D}z \sqrt{G} \: {\rm tr}
 \left(\frac{1}{4}F_{MN}F^{MN}\right).\label{S_P}
\end{align}
where $D=\dim P$ and $F_{MN}=\partial_M A_N-\partial_N A_M +i[A_M,A_N]$.
In order to make the reduction, we perform the vertical-horizontal
decompostion for the gauge field $A_M(z)$ and
the derivatives $\partial_M$
according to (\ref{general vielbeins}). The gauge field is decomposed as 
\begin{align}
 &A_M(z)=A_\alpha(z)E^\alpha_M(x)+A_a(z)E^a_M(z).
\end{align}
After the reduction, horizontal components $A_\alpha$ and vertical components $A_a$
of the gauge field will be naturally identified
with the gauge field and the higgs fields on the base space, 
respectively. 
The field strength in the local Lorentz frame is 
rewritten as follows:
\begin{align}
 &F_{\alpha\beta}
 =\nabla^{(M)}_{\alpha} A_\beta-\nabla^{(M)}_\beta A_\alpha
 +i[A_\alpha,A_\beta]-b^a_{\alpha\beta}A_a
 +ib^a_\alpha \cL_aA_\beta-ib^a_\beta \cL_aA_\alpha,
 \notag\\
 &F_{\alpha a}
 =e^\mu_{\alpha}\partial_{\mu} A_a+i\:[A_\alpha,A_b]
 -f^{abc}b^b_\alpha A_c
 -i\cL_a A_\alpha
 +i b^b_\alpha \cL_bA_a, \notag\\
 &F_{ab}
 =f^{abc}A_c+i\:[A_a,A_b]+i\cL_a A_b-i\cL_b A_a.
 \label{F on P}
\end{align}
Here we have defined the following quantities:
\begin{align}
 b^a_{\alpha\beta}&\equiv e^\mu_\alpha e^\nu_\beta\{\partial_\mu
 b^a_\nu-\partial_\nu b^a_\mu-f^{abc}b^b_\mu b^c_\nu\}, \notag\\
 \nabla^{(M)}_{\alpha}A_\beta
 &\equiv 
 e^\mu_\alpha\left(\partial_\mu A_\beta
 +\omega_{\mu\: \beta}^{\quad \gamma}A_\gamma\right), \notag\\
 \cL_a&\equiv -ie^m_a\partial_m,
\end{align}
where $\omega$ is the spin connection on the base space defined by 
$e^\alpha_\mu$ and $\cL_a$ are the right invariant Killing vectors on
the total space,
which represent the isometry.
Note that our calculations have been performed on $U$ so far.
When it is performed on $U'$, the quantities on $U'$ must be used.
The transformation of $b^a_\alpha(x)$ between on $U$ and on $U'$
is given by (\ref{b' and b}), so that that of $b^a_{\alpha\beta}(x)$ is
given by
\begin{align}
b'^a_{\alpha\beta}(x')=\Ad(k)^{ab}b^b_{\alpha\beta}(x).
\end{align}
The gauge field with the local Lorentz index
must be transformed according to (\ref{transf. of E}) as
\begin{align}
 &A'_\alpha=A_\alpha, \notag\\
 &A'_a=\Ad(k)^{ab}A_b. \label{transf. of A}
\end{align}

In order to make the dimensional reduction, 
we relate the fields on the total space
to those on the base space as
\begin{align}
 A_\alpha &= a_\alpha,\notag\\
 A_a &=  \phi_a, \label{A and a}
\end{align}
where $a_\alpha$ are the gauge field in the local Lorentz frame 
and $\phi_a$ are higgs fields on the base space.
We assume 
the both sides in (\ref{A and a}) are independent of $y^m$.
Using subscript of curved space, we can write (\ref{A and a})
equivalently as
\begin{align}
 A_\mu&= a_\mu-b^a_\mu\phi_a, \notag\\
 A_m&= e^a_m\phi_a.
 \label{A and a in curved space notation}
\end{align}
Here (\ref{A and a}) and (\ref{A and a in curved space notation})
are a generalization of 
(\ref{gauge fields on S^3 and gauge fields and higgs field on S^2})
and 
(\ref{gauge fields on S^3 and gauge fields and higgs field on S^2 2}),
respectively.
Substituting (\ref{F on P}) and (\ref{A and a}) into (\ref{S_P})
and using $\sqrt{G}=\sqrt{g}\sqrt{h}$,
we obtain YM-higgs on the base space:
\begin{align}
 S_M&=\frac{1}{g_M^2}\int d^d x \sqrt{g}\:{\rm tr}
 \bigg\{
 \frac{1}{4}\left(f_{\alpha\beta}-b^a_{\alpha\beta}\phi_a\right)^2
 +\frac{1}{2}\left(\nabla^{(M)}_\alpha \phi_a+i\:[a_\alpha,\phi_a]
 -f^{abc}b^{b}_\alpha\phi_c\right)^2 \notag\\
 &\qquad
 +\frac{1}{4}\left(f^{abc}\phi_c+i\:[\phi_a,\phi_b]\right)^2
 \bigg\},\label{S_M}
\end{align}
where $g_M^2=(\int dy \sqrt{h} )^{-1}g_{P}^2
=\frac{1}{\text{Vol}(G)}g_P^2$, $d=\dim M$ and $f_{\alpha\beta}
=\nabla^{(M)}_\alpha a_\beta-\nabla^{(M)}_\beta a_{\alpha}
+i[a_\alpha,a_\beta]$. Note that the connection in the fiber bundle can generate nontrivial mass terms of the higgs fields. This is reminiscent of the flux compactification in string theory.

\section{Extension of the matrix T-duality}
\setcounter{equation}{0} 
In this section, we extend the matrix T-duality on nontrivial $U(1)$  bundles developed in \cite{IIST}
to that on nontrivial $SU(2)$ bundles.

\subsection{Nontrivial vacua and transformation between patches}
As in the example in section 2, the theory on the base space (\ref{S_M})
has monopolelike vacua, which are in general patch-dependent if the principal bundle we consider is nontrivial.
Here we describe the vacua and their patch-dependence.
We also consider how the fields of the theory are transformed 
from a patch to another. We examine, in particular,
the transformation properties of fluctuations
around the vacua.

It is seen from (\ref{S_M}) that the condition for the vacua is given by
\begin{align}
& f_{\alpha\beta}-b^a_{\alpha\beta}\phi_a =0, \notag\\
& \nabla^{(M)}_\alpha \phi_a+i\:[a_\alpha,\phi_a]
 -f^{abc} b^{b}_\alpha\phi_c =0, \notag\\
& f^{abc}\phi_c+i\:[\phi_a,\phi_b] =0.
\end{align}
They are satisfied by the following configurations:
\begin{align}
 &\hat{a}_\alpha(x)=b^a_\alpha(x)\hat{\phi}_a=b^a_\alpha(x)L_a, \notag\\
 &\hat{\phi}_a=L_a, \label{vacua on M}
\end{align}
where $L_a$ are the generators of the Lie algebra of $G$
satisfying $[L_a,L_b]=if^{abc}L^c$ and generally reducible.
Note that as mentioned in section 3.1, $b^a(x)$ are generally patch-dependent
quantities. The vacua are, therefore, also patch-dependent.
From (\ref{b' and b}) and (\ref{vacua on M}), we can read off the
transformation properties for the vacua between patches:
\begin{align}
 &\hat{a}'(x)=K(x)\hat{a}(x)K(x)^{-1}+i\:dK(x)K(x)^{-1}, \notag\\
 &\hat{\phi}'_a=\Ad(k(x))_{ab}K(x)\hat{\phi}_bK(x)^{-1}=\hat{\phi}_a,
  \label{transf. of vacua btw patches}
\end{align}
where $K(x)$ is obtained by replacing $T^a$ in $k(x)$ in (\ref{transf. of g})
by $\hat{\phi}_a=L_a$.
Note that this is the gauge transformation by $K(x)$ 
except for the rotation of $\hat{\phi}_a$ by $\Ad(k(x))$,
which comes from (\ref{transf. of A}).

Let us consider the theory around the vacua (\ref{vacua on M}) and 
decompose the fields into the backgrounds and fluctuations:
\begin{align}
 &a_\alpha(x)=\hat{a}_\alpha(x)+\tilde{a}_\alpha(x), \notag\\
 &\phi_a(x)=\hat{\phi}_a+\tilde{\phi}_a(x).
\end{align}
The fluctuations are transformed between patches
as
\begin{align}
 & \tilde{a}'_\alpha(x)=K(x)\tilde{a}_\alpha(x) K(x)^{-1}, \notag\\
 & \tilde{\phi}'_a(x)=\Ad(k(x))_{ab}K(x)\tilde{\phi}_b(x)K(x)^{-1}.
 \label{transf. of fluctuations btw patches}
\end{align}
One can easily see that the action (\ref{S_M}) is indeed invariant under
the transformation (\ref{transf. of vacua btw patches}) and
(\ref{transf. of fluctuations btw patches}).

\subsection{$G=U(1),SU(2)$}
In this subsection, we consider the case in which the fiber is $U(1)$ or
$SU(2)$. In the case of $G=U(1)$, the matrix T-duality indeed works
as shown in \cite{IIST} and its typical example was given in section 2. 
We extend the matrix T-duality to the case of $G=SU(2)$
by applying the fact described in section 2 that YM on $S^3$ is realized
in the matrix model.

First, we review the matrix T-duality in the case of $G=U(1)$,
which is a generalization of the relationship between YM on $S^3$ and YM-higgs on $S^2$ in section 2.
In this case, the metric (\ref{metric}) reduces to the following form:
\begin{align}
 ds^2=g_{\mu\nu}(x)dx^\mu dx^\nu + (dy-b_\mu(x) dx^\mu)^2,
\label{metric of U(1) bundle}
\end{align}
where $y$ represents the fiber direction and $0\leq y < 2\pi$. We put $\dim M=d$.
(\ref{metric of S3}) indeed takes the form
of (\ref{metric of U(1) bundle}).
YM-higgs on the base space obtained from YM on the total space is given as the $U(1)$ case of
(\ref{S_M}):
\begin{align}
 S_M&=\frac{1}{g_M^2}\int d^d x \sqrt{g}\:{\rm tr}
 \bigg\{
 \frac{1}{4}\left(f_{\alpha\beta}-b_{\alpha\beta}\phi\right)^2
 +\frac{1}{2}\left(\nabla^{(M)}_\alpha \phi+i\:[a_\alpha,\phi]
 \right)^2 \bigg\}. \label{U(1) S_M}
\end{align}
(\ref{YM-higgs on S^2}) is a special case of (\ref{U(1) S_M}).
We show that we obtain the $U(N)$ YM on the total space from
the $U(N\times \infty)$ YM-higgs on the base space 
through the following procedure: we choose a certain background of 
the $U(N\times \infty)$ YM-higgs on the base space, 
expand the theory around the background and impose a periodicity 
condition. 

Note, first, that a general background of (\ref{U(1) S_M}) is given by
\begin{align}
 &\hat{a}_\alpha=b_\alpha\hat{\phi}, \notag\\
 &\hat{\phi}
 =-\:\mbox{diag}(\cdots,\underbrace{n_{s-1},\cdots,n_{s-1}}_{N_{s-1}},
 \underbrace{n_{s},\cdots,n_{s}}_{N_{s}},
 \underbrace{n_{s+1},\cdots,n_{s+1}}_{N_{s+1}},\cdots),
\end{align}
which is a counterpart of (\ref{vacuum of YM-higgs on S^2}).
We decompose the fields into the backgrounds and the fluctuations as
\begin{align}
 a_\alpha&\rightarrow \hat{a}_\alpha+a_\alpha, \notag\\
 \phi&\rightarrow \hat{\phi}+\phi. 
 \label{U(1) decomposition of base fields}
\end{align}
In particular, we take the following background:
$s$ running from $-\infty$ to $\infty$, $n_s=s$ and $N_s=N$. 
We label the (off-diagonal) blocks by $(s,t)$ and impose the periodicity
(orbifolding) condition on the fluctuations as in (\ref{periodicity condition}):
\begin{align}
 a_\alpha^{(s+1,t+1)}=a_\alpha^{(s,t)}\equiv a^{(s-t)}_\alpha, \notag\\
 \phi^{(s+1,t+1)}=\phi^{(s,t)}\equiv \phi^{(s-t)}.
\end{align}
The fluctuations are gauge-transformed from $U$ to $U'$ as
\begin{align}
 {a'}_\alpha^{(s-t)}&=e^{-i(s-t)v(x)}a^{(s-t)}_\alpha, \notag\\
 \phi'^{(s-t)}&=e^{-i(s-t)v(x)}\phi^{(s-t)},
 \label{U(1) transf. btw nhs}
\end{align}
where $e^{-iv}$ is a transition function; $e^{-iy'}=e^{-iv(x)}e^{-iy}$. 
(\ref{gauge transformation of fluctuations}) is a special case of (\ref{U(1) transf. btw nhs}). 
We make the Fourier transformation for the fluctuations on each patch to
construct the gauge field on the total space:
\begin{align}
 A_\alpha(x,y)
 &=\sum_{w}a^{(w)}_\alpha(x)e^{-iwy}, \notag\\
 A_{d+1}(x,y)
 &=\sum_{w}\phi^{(w)}(x)e^{-iwy}.
 \label{U(1) FT}
\end{align}
We can see from (\ref{U(1) transf. btw nhs}) that the lefthand sides 
in the above equations are indeed invariant under the transformation between patches.
Using (\ref{U(1) decomposition of base fields}) 
and (\ref{U(1) FT}), we can rewrite each term in (\ref{U(1) S_M}) as
\begin{align} 
 &\left(f_{\alpha\beta}-b_{\alpha\beta}\phi\right)^{(s,t)} \notag\\
 &\rightarrow
 \left(
 \nabla^{(M)}_\alpha a_\beta-\nabla^{(M)}_\beta a_\alpha
 +i[\hat{a}_\alpha,a_\beta]+i[a_\alpha,\hat{a}_\beta]+i[a_\alpha,a_\beta]
 -b_{\alpha\beta}\phi 
 \right)^{(s,t)} \notag\\
 &=
 \left(
 \nabla^{(M)}_\alpha a_\beta^{(s-t)}-\nabla^{(M)}_\beta a_\alpha^{(s-t)}
 +i[a_\alpha,a_\beta]^{(s-t)}
 -i(s-t)b_{\alpha}a_\beta^{(s-t)}
 +i(s-t)b_{\beta}a_\alpha^{(s-t)}
 -b_{\alpha\beta}\phi^{(s-t)}
 \right) \notag\\
 &=
 \frac{1}{2\pi}\int dy 
 \left(
 \nabla^{(M)}_\alpha A_\beta-\nabla^{(M)}_\beta A_\alpha
 +i[A_\alpha,A_\beta]-b_{\alpha\beta}A_{d+1}
 +b_\alpha \partial_y A_\beta -b_\beta \partial_y A_\alpha
 \right)\: e^{i(s-t)y} \notag\\
 &=\frac{1}{2\pi}\int dy F_{\alpha\beta}\: e^{i(s-t)y}, 
 \notag \\[5mm]
 &\left(\nabla^{(M)}_\alpha \phi+i\:[a_\alpha,\phi]\right)^{(s,t)}
 \notag\\
 &\rightarrow
 \left(\nabla^{(M)}_\alpha \phi+i\:[\hat{a}_\alpha,\phi]
 +i\:[a_\alpha,\hat{\phi}]+i\:[a_\alpha,\phi]\right)^{(s,t)} \notag\\
 &=\nabla^{(M)}_\alpha \phi^{(s-t)}+i\:[a_\alpha,\phi]^{(s-t)}
 -i\:(s-t)b_\alpha \phi^{(s-t)}
 +i\:(s-t)a_\alpha^{(s-t)}
 \notag\\
 &=
 \frac{1}{2\pi}\int dy\left(
 \nabla^{(M)}_\alpha A_{d+1}+i\:[A_\alpha,A_{d+1}]
 -\partial_y A_\alpha
 +b_\alpha \partial_y A_{d+1}\right)\:e^{i(s-t)y} \notag\\
 &=\frac{1}{2\pi}\int dy F_{\alpha (d+1)}\: e^{i(s-t)y}.
\label{matric T-duality for U(1)}
\end{align}
Then (\ref{U(1) S_M}) becomes
\begin{align}
 S_M
 &=\frac{1}{g_M^2}\int d^d x \sqrt{g}\:{\rm tr}
 \biggl\{
 \frac{1}{4}\left(f_{\alpha\beta}-b_{\alpha\beta}\phi\right)^2
 +\frac{1}{2}\left(\nabla^{(M)}_\alpha \phi_a+i\:[a_\alpha,\phi]
 \right)^2 \biggr\} \notag\\
 &=
 \frac{1}{g_M^2}\int d^d x \sqrt{g}\:{\rm tr}
 \Biggl[
 \sum_{s,t}
 \biggl\{
 \frac{1}{4}\left(f_{\alpha\beta}-b_{\alpha\beta}\phi\right)^{(s,t)}
 \left(f_{\alpha\beta}-b_{\alpha\beta}\phi\right)^{(t,s)} \notag\\
 &\hspace{5cm}
 +\frac{1}{2}
 \left(\nabla^{(M)}_\alpha \phi+i\:[a_\alpha,\phi]\right)^{(s,t)}
 \left(\nabla^{(M)}_\alpha \phi+i\:[a_\alpha,\phi]\right)^{(t,s)}
 \biggr\}
 \Biggr] \notag\\
 &\rightarrow
 \frac{1}{g_{M}^2}\frac{1}{2\pi}\sum_{w}\int d^D z \sqrt{G}\:
 \frac{1}{4}\tr\left(F_{AB}F_{AB}\right).
 \label{U(1) T-duality}
\end{align}
By dividing an overall factor $\sum_{w}$ in the last line in 
(\ref{U(1) T-duality}) to extract a single period, we obtain Yang-Mills
theory on the total space.

Next we consider the case where fiber is $SU(2)$.
In this case, YM-higgs on the base space takes the form
\begin{align}
 S_M&=\frac{1}{g_M^2}\int d^d x \sqrt{g}\:{\rm tr}
 \bigg\{
 \frac{1}{4}\left(f_{\alpha\beta}-b^a_{\alpha\beta}\phi_a\right)^2
 +\frac{1}{2}\left(\nabla^{(M)}_\alpha \phi_a+i\:[a_\alpha,\phi_a]
 -\epsilon^{abc} b^{b}_\alpha\phi_c\right)^2 \notag\\
 &\qquad\qquad
 +\frac{1}{4}\left(\epsilon^{abc}\phi_c+i\:[\phi_a,\phi_b]\right)^2
 \bigg\}.\label{SU(2) S_M}
\end{align}
We show that we can obtain the $U(N)$ YM on the total space
of a nontrivial $SU(2)$-bundle from the YM with three
higgs on its base space in a way similar to the case of $G=U(1)$.

The vacuum of YM-higgs is given by (\ref{vacua on M})
with $L_a$ satisfying the $SU(2)$ algebra, 
$[L_a,L_b]=i\: \epsilon_{abc} L_c$,
and $L_a$ generically take a reducible representation 
(\ref{matrix background}).
We expand the fields around this background,
\begin{align}
 a_\alpha(x)&\rightarrow \hat{a}_\alpha(x)+a_\alpha(x), \notag\\
 \phi_a(x)&\rightarrow \hat{\phi}_a+\phi_a(x).
 \label{SU(2) decomposition of base fields}
\end{align}
We label the (off-diagonal) blocks of the fluctuations by $(s,t)$,
which is $(N_s (2j_s+1))\times(N_t (2j_t+1))$ matrix,
and expand them by the fuzzy spherical harmonics:
\begin{align}
 a_\alpha^{(s,t)}(x)
 &=\sum_{J=|j_s-j_t|}^{j_s+j_t}\sum_{m=-J}^{J}
 a_{\alpha, Jm}^{(s,t)}(x) \otimes\hat{Y}_{Jm(j_sj_t)},
 \notag\\
 \phi_a^{(s,t)}(x)
 &=\sum_{J=|j_s-j_t|}^{j_s+j_t}\sum_{m=-J}^{J}
 \phi_{a, Jm}^{(s,t)}(x) \otimes\hat{Y}_{Jm(j_sj_t)}.
 \label{SU(2) fuzzy expansion}
\end{align}
We verify from (\ref{transf. of fluctuations btw patches}),
(\ref{SU(2) fuzzy expansion}) and (\ref{SU(2) algebra for fuzzy spherical harmonics}) that 
the modes are gauge-transformed from $U$ to $U'$ as
\begin{align}
 &a'^{(s,t)}_{\alpha, Jm}(x)
 =\sum_{m'}\langle Jm|k^{[J]}|Jm'\rangle a_{\alpha,Jm'}^{(s,t)}(x),
 \notag\\
 &\phi'^{(s,t)}_{a, Jm}(x)
 =\sum_{m'}
 \Ad(k)_{ab}\langle Jm|k^{[J]}|Jm'\rangle \phi_{b, Jm'}^{(s,t)}(x),
 \label{SU(2) transf. of the modes on the base}
\end{align}
where $k^{[J]}$ is the spin $J$ representation of $SU(2)$ for $k(x)$.

In what follows, we assume that as a background we set
$2j_s+1=N_0+s$ with $s$ running
from $-T$ to $T$ in (\ref{matrix background})
and take the limit of $N_0\rightarrow \infty$
and $T\rightarrow \infty$ in order.
For the modes, we impose the periodicity condition:
\begin{align}
 a_{\alpha,Jm}^{(s+1,t+1)}=a_{\alpha,Jm}^{(s,t)}
 \equiv a_{\alpha,Jm}^{(q_{st})}, \notag\\
 \phi_{a,Jm}^{(s+1,t+1)}=\phi_{a,Jm}^{(s,t)}
 \equiv \phi_{a,Jm}^{(q_{st})},
\end{align}
where $q_{st}\equiv \frac{s-t}{2}$. 
By using these modes and the spherical harmonics on $S^3$,
we make Fourier transformation on each patch 
to construct the gauge field on the total space:
\begin{align}
 &A_\alpha(z)
 =\sum_{Jm\tilde{m}} a_{\alpha,Jm}^{(\tilde{m})}(x)Y_{Jm\tilde{m}}(y),
 \notag\\
 &A_a(z)
 =\sum_{Jm\tilde{m}} \phi_{a, Jm}^{(\tilde{m})}(x)Y_{Jm\tilde{m}}(y).
 \label{SU(2) FT}
\end{align}
Its inverse is 
\begin{align}
 &a_{\alpha,Jm}^{(\tilde{m})}(x)
 =\int \frac{d\Omega_3}{2\pi^2} A_\alpha(z) Y_{Jm\tilde{m}}^\dagger (y),
 \notag\\
 &\phi_{a, Jm}^{(\tilde{m})}(x)
 =\int \frac{d\Omega_3}{2\pi^2} A_a(z) Y_{Jm\tilde{m}}^\dagger (y).
\label{SU(2) inverse FT}
\end{align}
From (\ref{SU(2) transf. of the modes on the base})
and (\ref{definition of spherical harmonics on S^3}),
it is verified that the lefthand sides in (\ref{SU(2) FT}) 
are indeed transformed between patches
as the gauge field on the total space (\ref{transf. of A}).

Using (\ref{SU(2) fuzzy expansion}) and
(\ref{SU(2) inverse FT}),
we can obtain the following equalities:
\begin{align}
 [L_a,a_\alpha(x)]^{(s,t)}
 &=\int \frac{d\Omega_3}{2\pi^2} \left(\cL_a A_\alpha(z)\right)
 Y_{Jpq_{st}}^\dagger \otimes \hat{Y}_{Jp(j_sj_t)},
\notag \\
[\phi_a,\phi_b]^{(s,t)}
&=\int\frac{d\Omega_3}{2\pi^2}
 [A_a(z),A_b(z)]Y_{Jmq_{st}}^\dagger(y)\otimes \hat{Y}_{Jm(j_sj_t)}.
 \label{fields on base to those on total}
\end{align}
The derivation of the above equalities is given in appendix B.
Substituting these into (\ref{SU(2) S_M}), we obtain
\begin{align}
 S_M
 &=\frac{1}{g_M^2}\int d^d x \sqrt{g}\:{\rm tr}
 \bigg\{
 \frac{1}{4}\left(f_{\alpha\beta}-b^a_{\alpha\beta}\phi_a\right)^2
 +\frac{1}{2}\left(\nabla^{(M)}_\alpha \phi_a+i\:[a_\alpha,\phi_a]
 -\epsilon^{abc} b^{b}_\alpha\phi_c\right)^2 \notag\\
 &\qquad\qquad
 +\frac{1}{4}\left(\epsilon^{abc}\phi_c+i\:[\phi_a,\phi_b]\right)^2
 \bigg\} \notag\\
 &=\frac{1}{g_M^2}\int d^d x \sqrt{g}\:{\rm tr}
 \Biggl[ \sum_{s,t}
 \biggl\{
 \frac{1}{4}
 \left(f_{\alpha\beta}-b^a_{\alpha\beta}\phi_a\right)^{(s,t)}
 \left(f_{\alpha\beta}-b^a_{\alpha\beta}\phi_a\right)^{(t,s)}
 \notag\\
 &\qquad\qquad
 +\frac{1}{2}
 \left(\nabla^{(M)}_\alpha \phi_a+i\:[a_\alpha,\phi_a]
 -\epsilon^{abc} b^{b}_\alpha\phi_c\right)^{(s,t)}
 \left(\nabla^{(M)}_\alpha \phi_a+i\:[a_\alpha,\phi_a]
 -\epsilon^{abc} b^{b}_\alpha\phi_c\right)^{(t,s)}
 \notag\\
 &\qquad\qquad
 +\frac{1}{4}
 \left(\epsilon^{abc}\phi_c+i\:[\phi_a,\phi_b]\right)^{(s,t)}
 \left(\epsilon^{abc}\phi_c+i\:[\phi_a,\phi_b]\right)^{(t,s)}
 \biggr\}
 \Biggr] \notag\\
 &\rightarrow
 \frac{1}{g_M^2}\frac{N_0}{2\pi^2}\sum_{w}
 \int d^D z \sqrt{G} \notag\\
 &\qquad\qquad 
 \times \tr \bigg\{
 \frac{1}{4}\left(
 \nabla^{(M)}_{\alpha} A_\beta-\nabla^{(M)}_\beta A_\alpha
 +i[A_\alpha,A_\beta]-b^a_{\alpha\beta}A_a
 +i b^a_\alpha \cL_aA_\beta-i b^a_\beta \cL_aA_\alpha
 \right)^2 \notag\\
 &\qquad\qquad 
 +\frac{1}{2}\left(
 \nabla^{(M)}_\alpha A_a+i\:[A_\alpha,A_b]
 -f^{abc} b^b_\alpha A_c
 -i\cL_a A_\alpha
 +i b^b_\alpha \cL_bA_a
 \right)^2 \notag\\
 &\qquad\qquad 
 +\frac{1}{4}\left(
 f^{abc}A_c+i\:[A_a,A_b]+i\cL_a A_b-i\cL_b A_a
 \right)^2
 \bigg\} \notag\\
 &= \frac{1}{g_M^2}\frac{N_0}{2\pi^2}\sum_{w}
 \int d^{d+1} z \sqrt{G}
 \mbox{tr}\left(\frac{1}{4}F_{AB}F_{AB}\right)
 \label{SU(2) S_M -> S_P}
\end{align}
By dividing an overall factor $\sum_{w}$ in the last line in 
(\ref{SU(2) S_M -> S_P}) to extract a single period, we obtain Yang-Mills
theory on the total space.

We can easily extend the above matrix T-duality to the case in which the fiber is $SU(2)^k\times U(1)^l$.
As an example, we consider an $SU(2)\times U(1)$ bundle, $P$. Let $a,b,c$ in (\ref{S_M}) run $0,1,2,3$
such that `$0$'
corresponds to the $U(1)$ direction and `$1,2,3$' correspond to the $SU(2)$ direction. We assign $i,j,k$ to
the $SU(2)$ direction.
We can consider YM-higgs on the $U(1)$ bundle on $M$, $M'$, which is obtained by making the dimensional reduction of the $SU(2)$ fiber direction for YM on the $SU(2)\times U(1)$ bundle.
We realize the theory
around an $SU(2)$ multimonopole background of YM-higgs on
$M'$ by taking the following background in YM-higgs on $M$ (\ref{S_M}) and imposing the periodicity condition
to the fluctuations:
\begin{align}
&\hat{\phi}_0=-\frac{1}{R}\mbox{diag}(\cdots,t-1,t,t+1,\cdots)\otimes 1_{\hat{M}}+b_0^i\hat{\phi}_i, \n
&\hat{\phi}_i=1_\infty\otimes (L_i \; \mbox{in (\ref{matrix background})}), \n
&\hat{a}_{\alpha}=b_{\alpha}^a\hat{\phi}_a,
\end{align}
where $b_{\alpha}^0$ represents the $U(1)$ monopole and $b_{\alpha}^i$ represents the SU(2) monopole.
$R$ is a certain constant depending on the fiber structure.
By setting  $2j_s+1=N_0+s$ with $s$ running
from $-T$ to $T$,
taking the limit of $N_0\rightarrow \infty$
and $T\rightarrow \infty$ in order and imposing the periodicity condition to the fluctuations again, 
we realize YM on $P$ in YM-higgs on $M$. In a similar way, we can realize YM on an $SU(2)^k\times U(1)^l$ in
YM-higgs on its base space.

\subsection{Example: $S^7\rightarrow S^4$}
We present an example of our findings in the previous subsection:
we consider $S^7$ with radius $2$ and regard it as $SU(2)\cong S^3$ Hopf
bundle on $S^4$ with radius $1$.

In order to describe $S^7$ as $SU(2)$ bundle on $S^4$,
it is convenient to introduce the quaternion $H$ (see for example \cite{Nakahara:2003nw,Naber:1997yu,Naber:2000bp}).
The quaternion algebra is defined by
\begin{align}
&\bi^2=\bj^2=\bk^2=-1,\quad  \bi\bj=-\bj\bi=\bk, \\
&\bj\bk=-\bk\bj=\bi,\quad  \bk\bi=-\bi\bk=\bj.
\end{align}
An arbitrary element of $H$ is written as
\begin{align}
q=a+b \bi +c \bj +d \bk.
\end{align}
where $a,b,c,d\in R$.
Its conjugation $q^*$ is defined by
\begin{align}
q^*\equiv a-b \bi -c \bj -d \bk.
\end{align}
The absolute value is given by
\begin{align}
 |q|\equiv \sqrt{q^*q}=\sqrt{a^2+b^2+c^2+d^2} \ge 0.
\end{align}

$S^7$ with radius $2$ is expressed by using quaternions as follows:
\begin{align}
 \{(q_1,q_2)\in H^2| |q_1|^2+|q_2|^2=4\}.
\end{align}
The Hopf map $\pi: S^7\rightarrow S^4$ is defined by
\begin{align}
 \pi: (q_1,q_2)\rightarrow [(q_1,q_2)]
\equiv \{(q_1,q_2)q|q\in H\backslash \{0\}\}.
\end{align}
In order to introduce local coordinates one needs to divide
$S^4$ in two patches: $U_1$ ($q_1\neq 0$) and $U_2$ ($q_2\neq 0$).
The local trivialization is given on each patch by
\begin{align}
 \pi^{-1}(U_1)\ni (q_1,q_2)
 &\rightarrow (q_2 q_1^{-1}, q_1|q_1|^{-1}) \in U_1\times SU(2),
 \notag\\
 \pi^{-1}(U_2)\ni (q_1,q_2)
 &\rightarrow (q_1 q_2^{-1}, q_2 |q_2|^{-1}) \in U_2\times SU(2).
 \label{local trivialization of S^7}
\end{align}
We parameterize $(q_1,q_2)$ by using a matrix representation of
quaternions as
\begin{align}
 q_1&=2\cos\frac{\chi}{2}\; \lambda, \notag\\
 q_2&=2\sin\frac{\chi}{2}\; \kappa \lambda.
 \label{explicit forms of lambda and kappa}
\end{align}
where $\kappa,\: \lambda\in SU(2)$ are defined 
by using Pauli matrices $\sigma^a\:(a=1,2,3)$ as
\begin{align}
 \kappa
 &=e^{i\eta\frac{\sigma^3}{2}}
 e^{i\xi\frac{\sigma^2}{2}}
 e^{i\zeta\frac{\sigma^3}{2}}, \notag\\
\lambda
 &=e^{-i\psi\frac{\sigma^3}{2}}
 e^{-i\theta\frac{\sigma^2}{2}}
 e^{-i\phi\frac{\sigma^3}{2}}.
\end{align}
The ranges of variables in the above equations are 
\begin{align}
 &0\leq \chi \leq \pi, \notag\\
 &0\leq \xi \leq \pi, \; 0\leq \eta < 2\pi, \; 0\leq \zeta <4\pi,
 \notag\\
 &0\leq \theta \leq \pi, \; 0\leq \phi < 2\pi, \; 0\leq \psi <4\pi.
\end{align}
In particular, $|\lambda|^2=\det\lambda=1$ and $|\kappa|^2=\det\kappa=1$ hold.
One can easily see from (\ref{local trivialization of S^7}) 
and (\ref{explicit forms of lambda and kappa})
that on $U_1$ the fiber space $SU(2)$ is described by $\lambda$ while
on $U_2$ that is described by $\lambda'\equiv \kappa \lambda$.
In the following, we restrict ourselves to the region $U_1$.
We denote sets of coordinates as 
$x^\mu=(\chi,\xi,\eta,\zeta)=(\chi, x^{\bar{\mu}})$ and
$y^m=(\theta, \phi, \psi)$. $x^{\mu}$ are 
coordinates of $S^4$, $x^{\bar{\mu}}$ are those of $S^3$
inside of $S^4$ and $y^m$ are those of $SU(2)$ of fiber. 
In order to describe a metric of $S^7$ explicitly,
we introduce the Maurer-Cartan 1-forms for $\kappa$ and $\lambda$
\begin{align}
 \kappa(\bar{x})^\dagger d\kappa(\bar{x})
 &= i\bar{e}^a_{\bar{\mu}}(\bar{x})\frac{\sigma^a}{2} dx^{\bar{\mu}},
 \notag\\
 d\lambda(y) \lambda(y)^\dagger 
 &= -i\: e^a_m(y)\frac{\sigma^a}{2} dy^m, \label{1-form of kappa and lambda}
\end{align}
where $\bar{x}$ represents the set of $\{x^{\bar{\mu}}\}$.
Then we define the metric of $S^7$ as
\begin{align}
 ds_{S^7}^2
 &=\det(dq_1)+\det(dq_2),
\end{align}
which is evaluated as
\begin{align}
ds_{S^7}^2=\Bigl(
 d\chi^2
 + \frac{1}{4} \sin^2\chi 
 \;\bar{e}^a_{\bar{\mu}}(\bar{x})\bar{e}^a_{\bar{\nu}}(\bar{x})
 dx^{\bar{\mu}} dx^{\bar{\nu}}
 \Bigr)
 +
 \Bigl(
 e^a_m(y)dy^m
 -\sin^2\frac{\chi}{2}\: \bar{e}^a_{\bar{\mu}}(\bar{x}) dx^{\bar{\mu}}
 \Bigr)^2.
 \label{metric of S^7}
\end{align}
In the above expression, the first term represents the metric of
the base space $S^4$ and the second one represents that of the fiber
space $SU(2)$ locally. 
Note that $\frac{1}{4}\bar{e}^a_{\bar{\mu}}\bar{e}^a_{\bar{\nu}}$
and $\frac{1}{4}e^a_{m}e^a_{n}$ are a metric of $S^3$ with radius $1$.
From (\ref{metric of S^7}) one can read off the vielbein on $S^4$ and
the local connections of the fiber bundle as
\begin{align}
 e^{\alpha}_{\;\; \mu}(x)&=
 \begin{pmatrix}
  1 & 0 \\
  0 & \frac{1}{2}\sin\chi \: \bar{e}^{a}_{\;\; \bar{\mu}}(\bar{x})
 \end{pmatrix}, 
 \qquad
 e^{\mu}_{\;\; \alpha}(x)=
 \begin{pmatrix}
  1 & 0 \\
  0 & \frac{2}{\sin\chi} \: \bar{e}^{\bar{\mu}}_{\;\; a}(\bar{x})
 \end{pmatrix}, 
 \notag\\
 &b^a_\chi(x)=0, 
 \quad 
 b^a_{\bar{\mu}}(x)
 =\tan\frac{\chi}{2}\: e^{a}_{\bar{\mu}}(x), \notag\\
 &b^a_{\chi\bar{\nu}}(x)
 =e^a_{\bar{\nu}}(x), 
\quad b^a_{\bar{\mu}\bar{\nu}}(x)
 =f^{abc} e^b_{\bar{\mu}}(x)e^c_{\bar{\nu}}(x).
 \label{S^7 connection}
\end{align}
As noted before, when we move to the other region, $U_2$, 
we must change $\lambda$ to $\lambda'\equiv \kappa \lambda$.
Then, one can easily find that the local connections change to
\begin{align}
 &{b'}^a_\chi(x)=0,
\quad
 {b'}^a_{\bar{\mu}}(x)
 =-\cot\frac{\chi}{2}\: \Ad(\kappa)^{ab}e^{b}_{\bar{\mu}}(x),
 \notag\\
 &b^a_{\chi\bar{\nu}}(x)
 =\Ad(\kappa)^{ab}b^b_{\chi\bar{\nu}}(x),
\quad
 b^a_{\bar{\mu}\bar{\nu}}(x)
 =\Ad(\kappa)^{ab}b^b_{\bar{\mu}\bar{\nu}}(x).
 \label{S^7 connection'}
\end{align}
This transformation property is consistent with (\ref{b' and b}).
The vacua of (\ref{SU(2) S_M}) are given by (\ref{vacua on M}), (\ref{matrix background}), 
(\ref{S^7 connection}) and (\ref{S^7 connection'}) on each patch.
$b^a_{\mu}$ and ${b'}^a_{\mu}$ are known as the gauge field of 
the Yang monopole \cite{Yang:1977qv}.

By applying the arguments in the previous subsection, we can show that
YM on $S^7$ is equivalent to the theory around the multi Yang monopole background of YM-higgs on $S^4$
with the periodicity imposed. 

\section{Gauge theories on $SU(n+1)(/H)$ and matrix model}
\setcounter{equation}{0}
In this section, we reveal various relations among 
gauge theories on $SU(n+1)$ and $SU(n+1)/H$,
where $H$ is $SU(n)$ or $SU(n)\times U(1)$ or $SU(n+1)$ which is a subgroup of $SU(n+1)$.
Note that $SU(n+1)/SU(n)\simeq S^{2n+1}$ and $SU(n+1)/(SU(n)\times U(1))\simeq CP^n$ and 
for $H=SU(n+1)$ the corresponding gauge theory reduces to a matrix model.
First, we develop a general formalism of a dimensional reduction by which one 
can obtain YM-higgs on $\tilde{G}/H$
from YM on $\tilde{G}$, where $\tilde{G}$ is an arbitrary group manifold.
Applying this formalism
to the case of $\tilde{G}=SU(n+1)$, we obtain 
YM-higgs on $S^{2n+1}$ and on $CP^n$
and the matrix model.
Next, by using the facts explained in appendix E, we show that the YM-higgs  
on $CP^n$ in the most general $U(1)$ monopole background is obtained by taking 
the commutative limit of the theory around a certain background of
the matrix model. We have found the correct form of 
the YM-higgs type action of such theory on $CP^n$.
Third, by using the extended matrix T-duality of the $U(1)$ case reviewed in section 4,
we show that YM-higgs on $S^{2n+1}$ is equivalent to the theory around a certain background of 
YM-higgs on $CP^n$ with the orbifolding condition imposed.
Combining these two facts,
we also show that YM-higgs on $S^{2n+1}$ 
is realized as the theory around an appropriate background 
of the matrix model with the orbifolding condition imposed. 
Finally, by using the results in section 4,  we show that
YM on $SU(n+1)$ is realized in 
YM-higgs on $SU(N+1)/(SU(2)^k\times U(1)^l)$. In particular, 
it follows that YM on $SU(3)$ is realized in YM-higgs on $S^5$ and on $CP^2$.

\subsection{Dimensional reduction of YM theory on a group manifold}
\label{Dimensional reduction of YM theory on a group manifold}

In this subsection, we restrict ourselves to the case 
in which the total space $P$ is itself a group manifold $\tilde{G}$.
In this case, we can take the Maurer-Cartan basis and 
rewrite the YM action on $\tilde{G}$ 
in such a way that the relation between YM on the 
total space and YM-higgs on the base space becomes more manifest.
In terms of this expression of the YM action, 
we can easily perform the dimensional reduction to 
obtain the YM-higgs theory on 
a coset space $\tilde{G}/H$, where $H$ is a subgroup of $\tilde{G}$.
Some conventions on the group manifold $\tilde{G}$ and 
the coset space $\tilde{G}/H$ are summarized in appendix C.

Let us consider pure YM on $\tilde{G}$.
In the Maurer-Cartan basis, the gauge potential
is written as $A=X_AE^A$ where $E^A$ are
the right invariant 1-forms on $\tilde{G}$ which are 
defined in (\ref{left and right invariant 1-forms}). 
In this basis, the field strength is written as
\begin{eqnarray}
F&=&dA+iA\wedge A \nonumber\\
&=& \frac{1}{2}\left(
f_{ABC}X_C+i{\cal L}_AX_B-i{\cal L}_BX_A +i[X_A,X_B]
\right)
E^A \wedge E^B,
\end{eqnarray} 
where we have used the Maurer-Cartan equation
(\ref{Maurer-Cartan equation}) and ${\cal L_A}$ are 
the right invariant Killing vectors on $\tilde{G}$ which 
are defined in (\ref{Killing vector on group manifold}). 
This is a counterpart of (\ref{F in MC basis}).
Then, the original YM action on $\tilde{G}$ 
is rewritten as follows:
\begin{align}
\frac{1}{g_{\tilde{G}}^2}\int {\rm tr}\left(\frac{1}{2}F\wedge *F \right)
=\frac{1}{g_{\tilde{G}}^2}\int d^{D} z \sqrt{G}\;{\rm tr} \left\{
\frac{1}{4}
\left(
f_{ABC}X_C+i{\cal L}_AX_B-i{\cal L}_BX_A +i[X_A,X_B]
\right)^2 \right\},
\label{YM action in terms of Y}
\end{align}
where $D={\rm dim}(\tilde{G})$, $G=\det G_{MN}$ and 
$G_{MN}$ is the metric on $\tilde{G}$. 
Note that the gauge transformation in this basis is given by
\begin{align}
X_A \rightarrow UX_AU^{-1}-{\cal L}_AU\:U^{-1}.
\end{align}
As explained in appendix C, 
if one drops the derivatives along the fiber direction in 
${\cal L}_A$, these operators are reduced to the 
$L_A$ which are the Killing vectors on $\tilde{G}/H$ 
defined in (\ref{Killing vector on coset space}). 
By dropping the derivatives along the fiber direction in 
${\cal L}_A$ in (\ref{YM action in terms of Y}),
therefore, we can obtain the theory on $\tilde{G}/H$,
\begin{align}
\frac{1}{g^2_{\tilde{G}}}
\int {\rm tr}\left(\frac{1}{2}F\wedge *F \right)
\rightarrow
\frac{1}{g^2_{\tilde{G}/H}}
\int d^d x\sqrt{g} \;{\rm tr} \left\{\frac{1}{4}
\left( f_{ABC}X_C+iL_AX_B-iL_BX_A +i[X_A,X_B] \right)^2
\right\},
\label{action of G/H}
\end{align}
where $g^2_{\tilde{G}/H}=g^2_{\tilde{G}}/{\rm Vol}(H)$,
$d=\dim{\tilde{G}/H}$, $g=\det g_{\mu\nu}$ and $g_{\mu\nu}$ is the metric on 
$\tilde{G}/H$. This is a counterpart of (\ref{YM-higgs on S^2 2}).

The action (\ref{action of G/H}) is also rewritten into the 
YM-higgs form which was obtained in section 3. 
The relation between 
the fields $X_A$ and the gauge and higgs fields on
$\tilde{G}/H$ is given as follows.
We introduce the orthogonal vectors to $L_A$ as
\begin{equation}
N_A^a={\rm Ad}(L(x))_A{}^a,
\end{equation}
where $L(x)$ is a representative element of $\tilde{G}/H$
which is defined in (\ref{g=Lh}),
and ${\rm Ad}$ represents the adjoint action:
$gT^Ag^{-1}=T^B{\rm Ad}(g)_{BA}$.
One can show the orthonormality conditions,
\begin{equation}
L_A^{\mu}L_A^{\nu}=-g^{\mu\nu},\;\; N_A^aN_A^b=\delta^{ab}
,\;\; L_A^{\mu}N_A^a=0,
\label{orthonormality of K and N}
\end{equation}
where $g^{\mu\nu}$ is the inverse of the metric on $\tilde{G}/H$.
Furthermore, the following equalities hold:
\begin{eqnarray}
&&L_A^{\mu}\partial_{\mu}N_B^a-
L_B^{\mu}\partial_{\mu}N_A^a=
-2iL_A^{\mu}L_B^{\nu}b_{\mu\nu}^a
-f_{abc}(L_A^{\mu}N_B^b-L_B^{\mu}N_A^b)b_{\mu}^c,
\nonumber\\
&& 
f_{ABC}N_C^a-f_{abc}N_A^bN_B^c+L_A^{\mu}L_B^{\nu}b_{\mu\nu}^a=0.
\label{relations satisfied by L and N}
\end{eqnarray}
We decompose $X_A$ into the gauge and higgs fields
in terms of $L_A^{\mu}$ and $N_A^a$ as follows \cite{Kitazawa:2002xj}: 
\begin{equation}
X_A=iL_A^{\mu}a_{\mu}+N_A^a\phi_a.
\label{relation between X and a}
\end{equation}
This is a generalization  of (\ref{X vector}).
Then, each term in the action (\ref{action of G/H}) 
is rewritten as
\begin{eqnarray}
f_{ABC}X_C&=&if_{ABC}L_C^{\mu}a_{\mu}
            -L_A^{\mu}L_B^{\nu}b_{\mu\nu}^a\phi_a
            +f_{abc}N_A^bN_B^c\phi_a, \nonumber\\
iL_AX_B-iL_BX_A&=& -if_{ABC}L_C^{\mu}a_{\mu}
                   -L_A^{\mu}L_B^{\nu}(\partial_{\mu}a_{\nu}
                       -\partial_{\nu}a_{\mu}-2b_{\mu\nu}^a\phi_a)
                    \nonumber\\
               && +i(L_A^{\mu}N_B^a-L_B^{\mu}N_A^a)
                    (\partial_{\mu}\phi_a- f_{abc}b_{\mu}^b\phi_c),
                    \nonumber\\ 
i[X_A,X_B]&=& -iL_A^{\mu}L_B^{\nu}[a_{\mu}, a_{\nu}]
               -(L_A^{\mu}N_B^a-L_B^{\mu}N_A^a)[a_{\mu},\phi_a]
               +iN_A^aN_B^b[\phi_b,\phi_a],
\label{rewrite of YM-higgs}
\end{eqnarray} 
where we have used (\ref{relations satisfied by L and N}).
By substituting these equations into the action 
(\ref{action of G/H}) and using (\ref{orthonormality of K and N}),
we indeed obtain the YM-higgs type action (\ref{S_M}),
\begin{eqnarray}
S_{\tilde{G}/H}=\frac{1}{g^2_{\tilde{G}/H}}
\int d^d x \sqrt{g} \; {\rm tr} \Biggl\{
\frac{1}{4} \left( f_{\mu \nu }-b_{\mu \nu }^a \phi_a \right) ^2
+\frac{1}{2} \left( D_{\mu} \phi_a-f_{abc}b_{\mu}^b \phi_c \right) ^2
\nonumber\\
+\frac{1}{4}\left(f_{abc}\phi_c+i[\phi_a,\phi_b]\right)^2
\Biggr\}.
\label{YM-higgs action on G/H}
\end{eqnarray}

Finally, we consider the case in which $P=\tilde{G}$ and the base 
manifold is just a point. This is the special case of the above 
dimensional reduction in which $H$ equals $\tilde{G}$ itself.
In this case, the theory on the base space is given by 
a zero-dimensional matrix model.
Dropping all the derivatives in 
(\ref{YM action in terms of Y}), we can easily make a dimensional
reduction to the matrix model:
\begin{eqnarray}
\frac{1}{g^2_{\tilde{G}}}
\int {\rm tr}\left( \frac{1}{2}F\wedge *F \right)
\rightarrow
\frac{1}{g^2_{mm}}\;{\rm tr} \left\{
\frac{1}{4} \left( f_{ABC}X_C+i[X_A,X_B] \right)^2\right\},
\label{reduction to MM}
\end{eqnarray}
where $g^2_{mm}=g^2_{\tilde{G}}/{\rm Vol}(\tilde{G})$. This is a counterpart of (\ref{SU(2) matrix model}).
Of course, we can obtain the matrix model (\ref{reduction to MM})
also from the theory (\ref{action of G/H}) on $\tilde{G}/H$ by 
dropping the derivatives $L_A$.
If we regard the original YM on $\tilde{G}$ as YM on a principal $\tilde{G}$ bundle over a point,
we obtain (\ref{reduction to MM}) as a special case of (\ref{S_M}).

\subsection{Dimensional reduction of YM theory on $SU(n+1)$}
\label{Dimensional reduction of YM on $SU(n+1)$}
In this subsection, we derive the YM-higgs on $S^{2n+1}$
and on $CP^n$ by applying the dimensional reduction  
discussed in the previous subsection.
We also derive the 0-dimensional matrix model  in which 
the YM-higgs on $S^{2n+1}$ and on $CP^n$ will be realized.  

Let us consider the group manifold $SU(n+1)$. 
We can apply the dimensional reduction developed in section
\ref{Dimensional reduction of YM theory on a group manifold} 
to the case of $P=\tilde{G}=SU(n+1)$
and obtain a theory on a coset space $SU(n+1)/H$, where 
$H$ is a subgroup of $SU(n+1)$.
We begin with pure YM on the group manifold $SU(n+1)$
in the Maurer-Cartan basis,
\begin{equation}
S_{SU(n+1)}=\frac{1}{g^2_{SU(n+1)}}
\int d^{n(n+2)} z\sqrt{G}\;{\rm tr} \left\{\frac{1}{4}
\left( f_{ABC}X_C+i{\cal L}_AX_B-i{\cal L}_BX_A +i[X_A,X_B] \right)^2
\right\}, \label{theory on SU(n+1)}
\end{equation}
where $f_{ABC}$ is the structure 
constant of $SU(n+1)$, $G=\det G_{MN}$ and 
$G_{MN}$ is the Cartan-Killing metric on $\tilde{G}$ which is
defined in (\ref{Cartan-Killing metric}).

Let us consider the dimensional reduction of the above theory 
to a theory on $\tilde{G}/H$.
If we take $H$ to be $SU(n)$, the coset space is given by
$SU(n+1)/SU(n) \simeq S^{2n+1}$. By applying the 
dimensional reduction (\ref{action of G/H}) to YM on $SU(n+1)$,
therefore, we obtain the YM-higgs theory on $S^{2n+1}$,
\begin{equation}
S_{S^{2n+1}}=\frac{1}{g^2_{S^{2n+1}}}
\int d^{2n+1} \tilde{x}\sqrt{\tilde{g}} \;{\rm tr} \left\{\frac{1}{4}
\left( f_{ABC}X_C+i\tilde{L}_AX_B-i\tilde{L}_BX_A +i[X_A,X_B] \right)^2
\right\}, \label{theory on S^{2n+1}}
\end{equation}
where $\tilde{g}$ represents the determinant of the metric 
on $S^{2n+1}$, and $\tilde{L}_A$'s are the Killing vectors on 
$S^{2n+1}$. Note that $S^{2n+1}$ that we consider here 
possesses only $SU(n+1)$ isometry which is smaller than 
$SO(2n+2)$. 
In fact, this is not the ordinary round sphere 
but a squashed sphere. 
In the case of $n=2$, the metric of this squashed $S^{5}$
is explicitly given in appendix D.

Next, we consider the case of $H=SU(n)\times U(1)$. In this case,
the coset space is $SU(n+1)/(SU(n)\times U(1)) \simeq CP^n$. 
Then, we can obtain the theory on $CP^n$ from YM on $SU(n+1)$ 
through the dimensional reduction, 
\begin{equation}
S_{CP^n}=\frac{1}{g^2_{CP^n}}
\int d^{2n} x \sqrt{g} \;{\rm tr} \left\{\frac{1}{4}
\left( f_{ABC}X_C+iL_AX_B-iL_BX_A +i[X_A,X_B] \right)^2
\right\}. \label{theory on CP^n}
\end{equation}
As in the case of $S^{2n+1}$, $g=\det g_{\mu\nu}$,
$g_{\mu\nu}$ and $L_A$ represent the metric and the Killing vectors
on $CP^n$ respectively.
The theory (\ref{theory on CP^n}) can be obtained also 
from the theory (\ref{theory on S^{2n+1}})
by dropping the derivative along the extra $U(1)$ fiber direction.
We can also rewrite (\ref{theory on S^{2n+1}}) and 
(\ref{theory on CP^n}) into the YM-higgs type actions
as in (\ref{YM-higgs action on G/H}) 
by using the relation (\ref{relation between X and a}).
For example, (\ref{theory on CP^n}) is rewritten 
into (\ref{YM-higgs action on G/H}) 
with $\mu,\nu=1,\cdots,2n$ and $a,b,c=0,\cdots,n^2-1$.
Here, $a,b,c$ are indices of $SU(n)\times U(1)$ and 
$a=0$ corresponds to the $U(1)$ direction.

Finally, we consider the case in which $H$ is $SU(n+1)$ itself.
In this case, the coset space is just a point.
Then, we obtain the following matrix model 
by using (\ref{reduction to MM}):
\begin{equation}
S_{mm}=\frac{1}{g^2_{mm}}
\;{\rm tr} \left\{\frac{1}{4}
\left( f_{ABC}X_C+i[X_A,X_B] \right)^2
\right\}. \label{matrix model}
\end{equation}
This theory is used to realize the theories
(\ref{theory on S^{2n+1}}) and (\ref{theory on CP^n})
in the next subsection. For $n=1$, the dimensional reductions in this subsection are equivalent to those in section 2.

\subsection{Relations among gauge theories on $SU(n+1)/H$} 
In this subsection,
we show that the theory (\ref{theory on CP^n}) in a monopole background
can be realized by taking the commutative limit of the theory around a  
nontrivial background of (\ref{matrix model}).
Combining this construction and the matrix T-duality, 
we also show that the theory (\ref{theory on S^{2n+1}})
on $S^{2n+1}$ can be realized as the theory around a certain 
background of the matrix model with the orbifolding condition imposed. 
Furthermore, we apply the extended matrix T-duality developed in section 4
to YM-higgs on $SU(n+1)/(SU(2)^k\times U(1)^l)$ and show that YM on $SU(n+1)$ is equivalent
to the theory around a certain vacuum of YM-higgs on $SU(n+1)/(SU(2)^k\times U(1)^l)$ with the periodicity
condition imposed.
For $n=2$, we obtain YM on $SU(3)$ from YM-higgs on $S^5$ and on $CP^2$ through the extended matrix T-duality.

First, we review nontrivial backgrounds of 
the theory (\ref{theory on CP^n}) on $CP^n$ and 
the matrix model (\ref{matrix model}).
The theory on $CP^n$ has many nontrivial monopole vacua. In particular, 
we focus on the $U(1)$ monopole background. 
Recall that we have $n^2$ higgs fields $\phi_a$. 
In the $U(1)$ monopole background, only the higgs field along the 
$U(1)$ direction $\phi_0$ acquires its nonzero vacuum expectation value.
In the gauge where $\phi_0$ is diagonal, 
the vacuum configurations of the $U(1)$ monopole with
the gauge group $U(M)$ are given by
\begin{align}
\hat{a}_{\mu}&=b_{\mu}^0 \hat{\phi}_0, \nonumber\\
\hat{\phi}_0 &= -\frac{1}{\sqrt{2n(n+1)}}{\rm diag}
(\cdots,\underbrace{n_{s-1},\cdots,n_{s-1}}_{N_{s-1}},
\underbrace{n_{s},\cdots,n_{s}}_{N_{s}},\underbrace{n_{s+1},\cdots,n_{s+1}}_{N_{s+1}},\cdots), \nonumber\\
\hat{\phi}_a&=0,\;\; ({\rm for}\;\; a\neq 0).
\label{U(1) monopole on CP^n written by a}
\end{align}
Here, $\sum_{s}N_s=M$ and $n_s$ must be integers 
due to Dirac's quantization condition. Because of
(\ref{relation between X and a}), the vacuum configurations of $X_A$ are
equivalently given by
\begin{equation}
\hat{X}_A=-\frac{iL_A^{\mu}b_{\mu}^0+N_A^0}{\sqrt{2n(n+1)}}\;
{\rm diag}(\cdots,\underbrace{n_{s-1},\cdots,n_{s-1}}_{N_{s-1}},
\underbrace{n_{s},\cdots,n_{s}}_{N_{s}},\underbrace{n_{s+1},\cdots,n_{s+1}}_{N_{s+1}},\cdots).
\label{U(1) monopole on CP^n written by X}
\end{equation}
The theory around the background 
(\ref{U(1) monopole on CP^n written by X})
is obtained by expanding each block of the fields in (\ref{theory on CP^n})
as $X_A^{(s,t)} \rightarrow \hat{X}_A^{(s,t)} +X_A^{(s,t)}$.
Then, the following action is obtained,
\begin{align}
\frac{1}{g^2_{CP^n}}
\int d^{2n} x \sqrt{g} \sum_{s,t} \;{\rm tr} \Big\{\frac{1}{4}
&\left( f_{ABC}X^{(s,t)}_C
+iL_A^{(q_{st})}X_B^{(s,t)}-iL_B^{(q_{st})}X_A^{(s,t)} 
+i[X_A,X_B]^{(s,t)} \right) \nonumber\\
\times &\left( f_{ABD}X^{(t,s)}_D
+iL_A^{(q_{ts})}X_B^{(t,s)}-iL_B^{(q_{ts})}X_A^{(t,s)} 
+i[X_A,X_B]^{(t,s)} \right) \Big\},
\label{theory on CP^n in U(1) monopole background}
\end{align}
where $q_{st}=\frac{n_s-n_t}{2}$ and $L_A^{(q)}$ are the angular momentum 
operators in the presence of a monopole with the magnetic charge $q$, 
which take the form
\begin{equation}
L_A^{(q)}=L_A+\frac{2q}{\sqrt{2n(n+1)}}(iL_A^{\mu}b_{\mu}^0+N_A^0).
\end{equation}
These operators are the generalization of (\ref{Lq}) in the case of $S^2$.

The vacua of the theory (\ref{matrix model}) are determined by
\begin{equation}
[X_A,X_B]=if_{ABC}X_C.
\end{equation}
In addition to the trivial solution $X_A=0$, there are
nontrivial solutions which are given by the representation matrices
of the $SU(n+1)$ generators,
\begin{equation}
\hat{X}_A=\hat{L}_A.
\label{clasical solution of SU(n+1) matrix model}
\end{equation} 
$\hat{L}_A$ are generally in a reducible representation.
In order to construct a theory on $CP^n$ in a $U(1)$ 
monopole background, we consider the following representation:
\begin{align}
 \hat{L}_A=
 \begin{pmatrix}
  \rotatebox[origin=tl]{-35}
  {$\cdots \;\;\;
  \overbrace{\rotatebox[origin=c]{35}{$\hat{L}_{A}^{(s-1)}$} \;
  \cdots \;
  \rotatebox[origin=c]{35}{$\hat{L}_{A}^{(s-1)}$}}^{\rotatebox{35}{$N_{s-1}$}}
  \;\;\;
  \overbrace{\rotatebox[origin=c]{35}{$\hat{L}_A^{(s)}$} \;
  \cdots \;
  \rotatebox[origin=c]{35}{$\hat{L}_A^{(s)}$}}^{\rotatebox{35}{$N_s$}}
  \;\;\;
  \overbrace{\rotatebox[origin=c]{35}{$\hat{L}_{A}^{(s+1)}$} \;
  \cdots \;
  \rotatebox[origin=c]{35}{$\hat{L}_A^{(s+1)}$}}^{\rotatebox{35}{$N_{s+1}$}}
  \;\;\; \cdots$}
 \end{pmatrix}.
\label{reducible representation of SU(n+1)}
\end{align}
Here $\hat{L}_A^{(s)}$ are the abbreviations of 
$\hat{L}_A^{[\Lambda_s,0,\cdots,0]}$ which 
are the generators of $SU(n+1)$ in the
irreducible representation specified by
the Dynkin index of $SU(n+1)$, $[\Lambda_s,0,\cdots,0]$. 
We consider the matrix model (\ref{matrix model}) 
around the background (\ref{reducible representation of SU(n+1)}) 
by expanding the each block of the fields around the background
:$X_A^{(s,t)} \rightarrow \hat{X}_A^{(s,t)}+X_A^{(s,t)}$.
Then, the action takes the following form:
\begin{align}
S_{mm}=\frac{1}{g^2_{mm}}
\sum_{s,t} \;{\rm tr} \Big\{\frac{1}{4}
&\left( f_{ABC}X_C^{(s,t)} +i\hat{L}_A\circ X_B^{(s,t)}
-i\hat{L}_B\circ X_A^{(s,t)} +i[X_A,X_B]^{(s,t)} \right) \nonumber\\
\times &\left( f_{ABD}X_D^{(t,s)} +i\hat{L}_A\circ X_B^{(t,s)}
-i\hat{L}_B\circ X_A^{(t,s)} +i[X_A,X_B]^{(t,s)} \right)
\Big\}. 
\label{matrix model around L_A}
\end{align}
$\hat{L}_A \circ$ are defined as
\begin{equation}
\hat{L}_A \circ X_B^{(s,t)} \equiv
\hat{L}_A^{(s)}X_B^{(s,t)}-X_B^{(s,t)}\hat{L}_A^{(t)}.
\end{equation}

We show in the following that the theory 
(\ref{matrix model around L_A})
is equivalent to the theory 
(\ref{theory on CP^n in U(1) monopole background})
if we put $\Lambda_s=N_0+n_s$ and take
$N_0 \rightarrow \infty$ limit. 
In order to show this equivalence, we make a harmonic expansion 
\cite{Alexanian:2001qj,Balachandran:2001dd,Dolan:2006tx}.
As explained in appendix E,
the $(s,t)$ blocks $X_A^{(s,t)}$ in the matrix model 
are expanded by the basis of rectangular matrices 
(\ref{fuzzy monopole harmonics on CP^n}) as
\begin{equation}
X_A^{(s,t)}=\sum_{J=|q_{st}|}^{(\Lambda_s+\Lambda_t)/2}
X_A^{(s,t)}{}^{{\bm \beta}_{J+q_{st}}}{}_{{\bm \alpha}_{J-q_{st}}} \otimes
\hat{Y}^{(q_{st})}{}_{{\bm \beta}_{J+q_{st}}}{}^{{\bm \alpha}_{J-q_{st}}}.
\label{expansion of blocks}
\end{equation}
Then, the diagonal coherent map allows us to map the $(s,t)$ 
blocks to local sections of the 
monopole bundle on $CP^n$ with the charge $q_{st}$,
\begin{align}
X_A^{(s,t)}&=\sum_{J=q_{st}}^{\infty}
X_A^{(q_{st})}{}^{{\bm \beta}_{J+q_{st}}}{}_{{\bm \alpha}_{J-q_{st}}} \otimes
\hat{Y}^{(q_{st})}{}_{{\bm \beta}_{J+q_{st}}}{}^{{\bm \alpha}_{J-q_{st}}}
\nonumber\\
&\rightarrow
\sum_{J=q_{st}}^{\infty}
X_A^{(q_{st})}{}^{{\bm \beta}_{J+q_{st}}}{}_{{\bm \alpha}_{J-q_{st}}}
\tilde{Y}^{(q_{st})}{}_{{\bm \beta}_{J+q_{st}}}{}^{{\bm \alpha}_{J-q_{st}}}
(w,\bar{w})
=X_A^{CP(s,t)}(w,\bar{w}),
\label{map1}
\end{align}
where we have taken the commutative limit $N_0 \rightarrow \infty$ and
$\tilde{Y}^{(q_{st})}{}_{{\bm \beta}_{J+q_{st}}}{}^{{\bm \alpha}_{J-q_{st}}}$
are the basis of local sections of the $U(1)$ monopole bundle on $CP^n$
which are defined in (\ref{monopole harmonics on CP^n}).
Note that we have put the superscript $CP$ on the quantity
in the righthand side of the above equation 
in order to emphasis that the $X_A^{CP(s,t)}$ are the 
fields on $CP^n$ appearing in 
(\ref{theory on CP^n in U(1) monopole background}). 
Similarly, $\hat{L}_A\circ$ is mapped to $L_A^{(q)}$ as shown in 
(\ref{Angular momentum operators on $CP^n$})\footnote{In \cite{Balachandran:2001dd}, (\ref{map1}) and (\ref{map2}) 
are proven to the quadratic order in the fields for all $q$ and to all order for $q=0$. In this paper,
we assume that these are also valid to all order for all $q$.}:
\begin{equation}
\hat{L}_A \circ X^{(s,t)}_B
\rightarrow
L_A^{(q)}X^{CP(s,t)}_B(w,\bar{w}).
\label{map2}
\end{equation}
Using (\ref{map1}) and (\ref{map2}), 
we find that the matrix model (\ref{matrix model around L_A})
is equivalent to the theory 
(\ref{theory on CP^n in U(1) monopole background}) on $CP^n$
in the commutative limit $N_0 \rightarrow \infty$.

Next, we show that the theory around a certain vacuum of 
$U(M=N\times \infty)$ YM-higgs on $CP^n$ with a
periodicity condition imposed is equivalent to $U(N)$ YM-higgs 
on $S^{2n+1}$. This statement is nothing but the matrix T-duality.
As explained in section 4, therefore, 
we consider the appropriate vacuum which is given by 
(\ref{U(1) monopole on CP^n written by X})
(or equivalently (\ref{U(1) monopole on CP^n written by a}))
with $s$ running from $-\infty$ to $\infty$, $n_s=s$ and 
$N_s=N$. We expand the fields on $CP^n$ around the background as
\begin{align}
X_A \rightarrow \hat{X}_A +X_A,
\end{align}
and impose the periodicity (orbifolding) condition on the 
fluctuation, 
\begin{align}
X_A^{(s+1,t+1)}=X_A^{(s,t)}\equiv X_A^{(s-t)}.
\end{align}
Then, we define the gauge and higgs fields on $S^{2n+1}$
by the Fourier transforms of the fluctuations
on each local coordinate patch:
\begin{align}
X_A^S=\sum_{u}X_A^{CP(u)}e^{-iuy},
\label{Fourier transformation on CP^n}
\end{align}
where $y$ is a coordinate which parameterizes 
the fiber ($U(1)$) direction and satisfies
$0 \leq \tau \leq 2\pi$.
Here, the superscripts $S$ and $CP$ indicate that
$X_A^S$ and $X_A^{CP(w)}$ are the fields on $S^{2n+1}$ and
$CP^{n}$ respectively.
We substitute (\ref{Fourier transformation on CP^n})
into (\ref{theory on CP^n in U(1) monopole background}) 
and divide an overall factor $\sum_s$ to 
extract a single period. Then, we obtain $U(N)$ YM-higgs
on $S^{2n+1}$ written in the basis of $X_A$ 
(\ref{theory on S^{2n+1}}).

Combining the above matrix T-duality and 
the construction of 
(\ref{theory on CP^n in U(1) monopole background}) 
in terms of the matrix model, we find that the theory 
around (\ref{reducible representation of SU(n+1)}) of
the matrix model, where $s$ runs from $-\infty$ to $\infty$ and
$\Lambda_s=N_0+s$, is equivalent to $U(N)$ YM-higgs on $S^{2n+1}$
if we take the limit $N_0 \rightarrow \infty$, impose
the periodicity condition on the fluctuations, and 
finally divide the overall factor $\sum_s$. 

Finally, it is straightforward to apply the extended matrix 
T-duality to $SU(2)^k\times U(1)^l$ bundle on $SU(n+1)/(SU(2)^k\times U(1)^l)$ and
show that YM on $SU(n+1)$ is equivalent to the theory around a certain vacuum of YM-higgs on
$SU(n+1)/(SU(2)^k\times U(1)^l)$ with the periodicity condition imposed.

\section{Interpretation as Buscher's T-duality}
\setcounter{equation}{0}
In this section, let us see that the extended matrix T-duality of the $U(1)$ case, which
was obtained in \cite{IIST} and reviewed in section 4.2, is actually interpreted as the T-duality in Buscher's sense.
We put $\mbox{dim}M=p$. For $G=U(1)$, as in (\ref{metric of U(1) bundle}), 
the metric of the total space is
given by  
\begin{align}
ds^2=G_{MN}dz^Mdz^N=g_{\mu\nu}dx^{\mu}dx^{\nu}+(dy-b_{\mu}dx^{\mu})^2,
\end{align}
where $M,N=1,\cdots, p+1$ and $\mu,\nu=1,\cdots,p$.
We assume that the other fields such as the antisymmetric fields and the dilaton field are trivial.
Then, YM on the total space is viewed as the low energy effective theory for the D$p$-branes wrapped on the
total space\footnote{Here we ignore the transverse directions.}.
We make the T-duality transformation for the fiber direction to obtain a new geometry \cite{Buscher:1987sk}:
\begin{align}
&{ds'}^2=g_{\mu\nu}dx^{\mu}dx^{\nu}+dy^2, \nonumber\\
&{B'}_{\mu\nu}=0,\;\;\;{B'}_{\mu y}=-b_{\mu}.
\label{new geometry}
\end{align}
The D$p$-branes should be transformed to the D$(p-1)$-branes wrapped on the base space.
The D$(p-1)$-brane effective action on the new geometry (\ref{new geometry}) is given by
\begin{align}
S_{p-1}=\tau_{p-1}\int d^p\sigma e^{-\Phi}\sqrt{\det(\tilde{G}_{ab}+\tilde{B}_{ab}+2\pi\alpha'F_{ab})},
\label{D(p-1)-brane effective action}
\end{align}
where $\sigma^a\; (a=1,\cdots,p)$ parameterize the world volume of the D$(p-1)$-brane, and
$\tilde{G}_{ab}$ and $\tilde{B}_{ab}$ are the pullback of (\ref{new geometry}) on the world volume which 
is defined through
the embedding of world volume $z^M(\sigma)$ as 
\begin{align}
&\tilde{G}_{ab}=\frac{\partial z^M}{\partial \sigma^a}\frac{\partial z^N}{\partial \sigma^b}{G'}_{MN}, \nonumber\\
&\tilde{B}_{ab}=\frac{\partial z^M}{\partial \sigma^a}\frac{\partial z^N}{\partial \sigma^b}{B'}_{MN}.
\end{align}
In the static gauge $x^{\mu}(\sigma)=\sigma^{\mu}$ and $z^y(\sigma)=2\pi\alpha'\phi$, 
(\ref{D(p-1)-brane effective action}) reduces to
\begin{align}
S_{p-1}=\tau_{p-1}\int d^px \sqrt{\det(g_{\mu\nu}+(2\pi\alpha')^2\partial_{\mu}\phi\partial_{\nu}\phi
+2\pi\alpha'(F_{\mu\nu}+\partial_{\mu}\phi b_{\nu}-\partial_{\nu}\phi b_{\mu}))}.
\end{align}
Up to ${\cal O}({\alpha'}^3)$, this equals
\begin{align}
\frac{1}{g_{YM}^2}\int d^px \sqrt{g}\left(
\frac{1}{4}(F_{\alpha\beta}+\nabla_{\alpha}\phi b_{\beta}-\nabla_{\beta}\phi b_{\alpha})^2
+\frac{1}{2}(\nabla_{\alpha}\phi)^2\right),
\label{D(p-1)-brane effective theory}
\end{align}
where $g_{YM}^2=\frac{1}{4\pi\alpha^2\tau_{p-1}}$. 
If we redefine the gauge field as $a_{\alpha} \rightarrow a_{\alpha}+b_{\alpha}\phi$ and 
make non-abelianization, we obtain from (\ref{D(p-1)-brane effective theory})
\begin{align}
\frac{1}{g_{YM}^2}\int d^px \sqrt{g}\mbox{tr}\left(
\frac{1}{4}(F_{\alpha\beta} -b_{\alpha\beta}\phi)^2
+\frac{1}{2}(D_{\alpha}\phi)^2\right),
\label{D(p-1)-brane effective theory 2}
\end{align}
which indeed agrees with (\ref{S_M}) with $G=U(1)$.

\section*{Acknowledgements}
We would like to thank T. Higashi for discussions.
The work of G.I. is supported in part by the JSPS Research Fellowship 
for Young Scientists.
The work of A.T. is supported in part by Grant-in-Aid for Scientific
Research (No. 19540294) from the Ministry 
of Education, Culture, Sports, Science and Technology.

\appendix

\section{Spherical harmonics}
\setcounter{equation}{0}
\renewcommand{\theequation}{A.\arabic{equation}}
In this appendix, we review the spherical harmonics on $S^3$, the monopole harmonics on $S^2$ \cite{Wu:1976ge} and
the fuzzy spherical harmonics \cite{Grosse:1995jt,Dasgupta:2002hx,ISTT}. 
For more details, see \cite{ITT,IIST} and references therein.
\subsection*{A.1\hspace{0.5cm}Spherical harmonics on $S^3$}
We regard $S^3$ as the $SU(2)$ group manifold. We parameterize an element
of $SU(2)$ in terms of the Euler angles as
\begin{equation}
g=e^{-i\varphi J_3}e^{-i\theta J_2}e^{-i\psi J_3},
\end{equation}
where $J_A$ satisfy $[J_A,J_B]=i\epsilon_{ABC}J_C$ and 
$0\leq \theta\leq \pi$, $0\leq \varphi < 2\pi$, $0\leq \psi < 4\pi$.
The isometry of $S^3$ is $SO(4)=SU(2)\times SU(2)$, and these two
$SU(2)$'s act on $g$ from left and right, respectively. We construct the
right invariant 1-forms, 
\begin{equation}
dgg^{-1}=-i\mu E^A J_A,
\end{equation}
where the radius of $S^3$ is $2/\mu$. They are explicitly
given by 
\begin{eqnarray}
&&E^1=\frac{1}{\mu}(-\sin \varphi d\theta + \sin\theta\cos\varphi d\psi),\nonumber\\
&&E^2=\frac{1}{\mu}(\cos \varphi d\theta + \sin\theta\sin\varphi
 d\psi),\nonumber\\
&&E^3=\frac{1}{\mu}(d\varphi + \cos\theta d\psi),
\end{eqnarray}
and satisfy the Maure-Cartan equation
\begin{equation}
dE^A-\frac{\mu}{2}\epsilon_{ABC}E^B\wedge E^C=0.\label{Maure-Cartan}
\end{equation}
The metric is constructed from $E^A$ as
\begin{equation}
ds^2=E^AE^A=\frac{1}{\mu^2}\left(
d\theta^2+\sin^2\theta d\varphi^2 +(d\psi+\cos d\varphi)^2\right).
\end{equation}
The Killing vectors dual to $E^A$ are given by
\begin{equation}
{\cal{L}}_A=-\frac{i}{\mu}E^M_A\partial_M,
\end{equation}
where $E^M_A$ are inverse of $E^A_M$. The explicit form of the Killing
vectors are
\begin{eqnarray}
&&{\cal{L}}_1=-i\left(-\sin\varphi\partial_{\theta}-\cot\theta\cos\varphi\partial_{\varphi}+\frac{\cos\varphi}{\sin\theta}\partial_{\psi}\right),\nonumber\\
&&{\cal{L}}_2=-i\left(\cos\varphi\partial_{\theta}-\cot\theta\sin\varphi\partial_{\varphi}+\frac{\sin\varphi}{\sin\theta}\partial_{\psi}\right),\nonumber\\
&&{\cal{L}}_3=-i\partial_{\varphi}.\label{Killing vector}
\end{eqnarray}
Because of the Maure-Cartan equation (\ref{Maure-Cartan}), the Killing vectors satisfy the SU(2) algebra, $[{\cal{L}}_A,{\cal{L}}_B]=i\epsilon_{ABC}{\cal{L}}_C$.

The scalar spherical harmonics on $S^3$ are given by
\begin{equation}
Y_{Jm\tilde{m}}(\Omega_3)=(-1)^{J-\tilde{m}}\sqrt{2J+1}\langle
 J-\tilde{m}|g^{-1}|Jm \rangle.
\label{definition of spherical harmonics on S^3}
\end{equation}
These spherical harmonics form the basis of SU(2) algebra generated
by ${\cal{L}}_A$'s.
\begin{eqnarray}
{\cal{L}}^2Y_{Jm\tilde{m}}&=&J(J+1)Y_{Jm\tilde{m}},\nonumber\\
{\cal{L}}_{\pm}Y_{Jm\tilde{m}}&=&\sqrt{(J\mp m)(J \pm
 m+1)}Y_{Jm\pm 1\tilde{m}},\nonumber\\
{\cal{L}}_3Y_{Jm\tilde{m}}&=&mY_{Jm\tilde{m}}.
\label{SU(2) alg. of S3 harmonics}
\end{eqnarray}
The complex conjugates of the spherical harmonics are evaluated as
\begin{equation}
\left(Y_{Jm\tilde{m}}\right)^*=(-1)^{m-\tilde{m}}Y_{J-m-\tilde{m}}.
\end{equation}
The spherical harmonics also satisfy the orthonormality condition
\begin{equation}
\int \frac{d\Omega_3}{2\pi^2}\left(Y_{Jm\tilde{m}}\right)^*
Y_{J'm'\tilde{m}'}=\delta_{JJ'}\delta_{mm'}\delta_{\tilde{m}\tilde{m}'}.
\end{equation}
The integral of the product of three spherical harmonics is
given as follows:
\begin{eqnarray}
{\cal{C}}^{J_1m_1\tilde{m}_1}_{J_2m_2\tilde{m}_2 J_3m_3\tilde{m}_3}
 &\equiv&
\int \frac{d\Omega_3}{2\pi^2}\left(Y_{J_1m_1\tilde{m}_1}\right)^*
Y_{J_2m_2\tilde{m}_2}Y_{J_3m_3\tilde{m}_3} \nonumber\\
&=&
\sqrt{\frac{(2J_2+1)(2J_3+1)}{2J_1+1}}
C^{J_1m_1}_{J_2m_2J_3m_3}C^{J_1\tilde{m}_1}_{J_2\tilde{m}_2J_3\tilde{m}_3},
\label{integral of three Ys}
\end{eqnarray}
where $C^{J_1m_1}_{J_2m_2J_3m_3}$ is the 
Clebsch-Gordan coefficient of $SU(2)$.
Finally, the spherical harmonics satisfy the 
completeness condition,
\begin{align}
\sum_{Jm\tilde{m}} \left( Y_{Jm \tilde{m}} \right)^* 
(\Omega _3) Y_{Jm \tilde{m}} (\Omega_3')
=2 \pi ^2 \delta ( \Omega_3-\Omega_3'),
\label{completeness condition of S3 harmonics}
\end{align}
where 
\begin{equation}
\delta (\Omega_3)=\frac{8}{\sin \theta} \delta(\theta)
\delta(\varphi) \delta (\psi).
\end{equation}

\subsection*{A.2\hspace{0.5cm}Monopole spherical harmonics on $S^2$}
We adopt the following metric for $S^2$:
\begin{equation}
ds^2=\frac{1}{\mu^2}(d\theta^2+\sin^2\theta d\varphi^2).
\end{equation}
We define two local patches on $S^2$ to describe nontrivial $U(1)$
bundles over $S^2$: the patch I is specified by $0\leq\theta <\pi$ and the
patch II is specified by  $0<\theta\leq\pi$. 
In the following expressions, the upper sign is taken in the patch I and
the lower sign in the patch II.

The angular momentum operator in the presence of a monopole with
magnetic charge $q$ at the origin takes the form
\begin{eqnarray}
&&L_1^{(q)}=i(\sin\varphi\partial_{\theta}+
\cot\theta\cos\varphi\partial_{\varphi})-
q\frac{1\mp \cos\theta}{\sin\theta}\cos\varphi, \nonumber\\
&&L_2^{(q)}=i(-\cos\varphi\partial_{\theta}+
\cot\theta\sin\varphi\partial_{\varphi})-
q\frac{1\mp \cos\theta}{\sin\theta}\sin\varphi, \nonumber\\
&&L_3^{(q)}=-i\partial_{\varphi}\mp q ,
\label{monopole angular momentum}
\end{eqnarray}
where $q$ is quantized as 
$q=0, \pm \frac{1}{2}, \pm 1, \pm \frac{3}{2},\cdots$.
These operators act on the local sections on $S^2$ and satisfy the $SU(2)$
algebra $[L_A^{(q)},L_B^{(q)}]=i\epsilon_{ABC}L_C^{(q)} $.
Note that when $q=0$, these operators are reduced to the ordinary
angular momentum operators on $S^2$ (or $R^3$).
  if we regard $S^3$ as a $U(1)$ bundle over $S^2$, 
and parameterize the fiber direction by $y=\psi\pm\varphi$,
the above expression (\ref{monopole angular momentum}) can be obtained
by making a replacement in (\ref{Killing vector}): 
$\partial_y \rightarrow -iq$.

The monopole spherical harmonics are the basis of local sections on
$S^2$ and also form the basis of the $SU(2)$ algebra generated by
$L_A^{(q)}$. The monopole scalar spherical harmonics are given by
\begin{equation}
\tilde{Y}_{Jmq}(\Omega_2)=
(-1)^{J-q}\sqrt{2J+1}\langle J -q|e^{i\theta J_2}|Jm\rangle
e^{i(\pm q+m)\varphi}.
\label{monopole harmonics}
\end{equation}
Here $J=|q|,|q|+1,|q+2|,\cdots$, $m=-J,-J+1,\cdots , J-1,J$.
The existence of the lower bound of the angular momentum $J$ 
is due to the fact that the magnetic field produced by the 
monopole also has nonzero angular momentum.  
Note that the monopole harmonics with $q=0$ do not transform
on the overlap of two patches. They correspond to 
global sections (functions) on $S^2$ which are expressed by
the ordinary spherical harmonics on $S^2$.
The action of $L^{(q)}_A$ on the monopole spherical harmonics 
is given by
\begin{eqnarray}
L^{(q)2}\tilde{Y}_{Jmq}&=&J(J+1)\tilde{Y}_{Jmq},\nonumber\\
L^{(q)}_{\pm}\tilde{Y}_{Jmq}&=&\sqrt{(J\mp m)(J \pm
 m+1)}\tilde{Y}_{Jm\pm 1q},\nonumber\\
L^{(q)}_3\tilde{Y}_{Jmq}&=&m\tilde{Y}_{Jmq}.
\end{eqnarray}
The complex conjugates of the monopole spherical harmonics 
are evaluated as
\begin{equation}
\left(\tilde{Y}_{Jmq}\right)^*=(-1)^{m-q}\tilde{Y}_{J-m-q}.
\end{equation}
The monopole spherical harmonics are orthonormal to each other,
\begin{equation}
\int \frac{d\Omega_2}{4\pi}\left(\tilde{Y}_{Jmq}\right)^*
\tilde{Y}_{J'm'q}=\delta_{JJ'}\delta_{mm'}.
\end{equation}
The integral of three monopole spherical harmonics is equal to 
the corresponding integral (\ref{integral of three Ys}) 
on $S^3$ with the identification $ \tilde{m}=q $,
\begin{eqnarray}
\int \frac{d\Omega_2}{4\pi}\left(\tilde{Y}_{J_1m_1q_1}\right)^*
\tilde{Y}_{J_2m_2q_2}\tilde{Y}_{J_3m_3q_3}
&=&
\sqrt{\frac{(2J_2+1)(2J_3+1)}{2J_1+1}}
C^{J_1m_1}_{J_2m_2J_3m_3}C^{J_1q_1}_{J_2q_2J_3q_3}
\nonumber\\
&=&
{\cal{C}}^{J_1m_1q_1}_{J_2m_2q_2 J_3m_3q_3},
\label{integral of three Y tilde}
\end{eqnarray}
where the monopole charges must be conserved 
in the lefthand side of the above equation as $q_1+q_2+q_3=0$. 
Note that the monopole spherical harmonics are expressed 
in terms of the spherical harmonics on $S^3$:
\begin{eqnarray}
\tilde{Y}_{Jmq}(\Omega_2)
&=&e^{iq(\psi \pm \varphi)}Y_{Jmq}(\Omega_3),\nonumber\\
L^{(q)}_A \tilde{Y}_{Jmq}(\Omega_2)
&=&e^{iq(\psi \pm \varphi)} {\cal L}_A Y_{Jmq}(\Omega_3).
\label{map between Y and Y tilde}
\end{eqnarray}
(\ref{integral of three Y tilde}) and 
(\ref{map between Y and Y tilde})
represent a map between the local sections on $S^2$ and the 
Kaluza-Klein modes on $S^3$.

\subsection*{A.3\hspace{0.5cm}Fuzzy spherical harmonics}
Let us consider $(2j+1)\times (2j'+1)$ rectangular complex matrices.
Such matrices are generally expressed as
\begin{equation}
M=\sum_{r,r'}M_{rr'}|jr\rangle\langle j'r'|.
\end{equation}
We can define linear maps $\hat{L}_A\circ$, which map the set of 
$(2j+1)\times (2j'+1)$ rectangular complex matrices to itself, by their 
operation on the basis:
\begin{equation}
\hat{L}_A \circ |jr\rangle\langle j'r'| \equiv 
\hat{L}_A^{[j]}|jr\rangle\langle j'r'|-
|jr\rangle\langle j'r'|\hat{L}_A^{[j']},
\end{equation}
where $\hat{L}_A^{[j]}$ are the spin $j$ representation matrices of
the $SU(2)$ generators. $\hat{L}_A \circ$ satisfy the 
$SU(2)$ algebra 
$[\hat{L}_A\circ ,\hat{L}_B\circ] = i\epsilon_{ABC}\hat{L}_C \circ$.

We make a change of a basis of the rectangular matrices 
from the above basis $\{|jr\rangle\langle j'r'|\}$
to the new basis which is called the fuzzy spherical harmonics:
\begin{equation}
\hat{Y}_{Jm(jj')}=\sqrt{N_0}\sum_{r,r'}(-1)^{-j+r'}C^{Jm}_{jr\;j'-r'}
|jr\rangle\langle j'r'|,
\label{fuzzy spherical harmonics}
\end{equation}
where $N_0$ is a positive integer which will be specified below.
For a fixed $J$ the fuzzy spherical harmonics also form a
basis of the spin $J$ irreducible representation of $SU(2)$ which is
generated by $\hat{L}_A \circ$,
\begin{eqnarray}
(\hat{L}_A \circ)^2\hat{Y}_{Jm(jj')}
&=&J(J+1)\hat{Y}_{Jm(jj')},\nonumber\\
\hat{L}_{\pm}\circ \hat{Y}_{Jm(jj')}
&=&\sqrt{(J\mp m)(J \pm
 m+1)}\hat{Y}_{Jm\pm 1(jj')},\nonumber\\
\hat{L}_3\circ \hat{Y}_{Jm(jj')}
&=&m\hat{Y}_{Jm(jj')}.
\label{SU(2) algebra for fuzzy spherical harmonics}
\end{eqnarray}
The hermitian conjugates of the fuzzy spherical harmonics 
are evaluated as
\begin{equation}
\left(\hat{Y}_{Jm(jj')}\right)^{\dagger}
=(-1)^{m-(j-j')}\hat{Y}_{J-m(j'j)}.
\end{equation}
The fuzzy spherical harmonics satisfy the orthonormality
condition under the following normalized trace:
\begin{equation}
\frac{1}{N_0}{\rm tr}\left\{
\left(\hat{Y}_{Jm(jj')}\right)^{\dagger}
\hat{Y}_{J'm'(jj')}\right\}=\delta_{JJ'}\delta_{mm'},
\end{equation}
where ${\rm tr}$ stands for the trace over 
$(2j'+1)\times(2j'+1)$ matrices. 
The trace of three fuzzy spherical harmonics is given by
\begin{align}
\hat{C}^{J_1m_1(jj'')}_{J_2m_2(jj')J_3m_3(j'j'')}
&\equiv
\frac{1}{N_0}{\rm tr} \left\{
\left(\hat{Y}_{J_1m_1(jj'')}\right)^{\dagger}
\hat{Y}_{J_2m_2(jj')}\hat{Y}_{J_3m_3(j'j'')}
\right\} \nonumber\\
&=
(-1)^{J_1+j+j''}\sqrt{N_0(2J_2+1)(2J_3+1)}
C^{J_1m_1}_{J_2m_2J_3m_3}
\left\{
\begin{array}{ccc}
J_1 & J_2 & J_3 \\
j' & j'' & j  
\end{array}
\right\},
\end{align} 
where the last factor of the above equation is the $6-j$ symbol.

In order to reveal relationships among the fuzzy spherical
harmonics, the monopole harmonics on $S^2$ and 
the spherical harmonics on $S^3$,
we introduce the following parameterization for $j$, $j'$ and $j''$,
\begin{equation}
2j+1=N_0+\zeta,\;\;2j'+1=N_0+\zeta',\;\;2j''+1=N_0+\zeta''.
\end{equation}
$\zeta$, $\zeta'$ and $\zeta''$ are integers which are grater 
than $-N_0$. 
Then, in the limit $N_0 \rightarrow \infty$, one can show that
\begin{equation}
\hat{C}^{J_1m_1(jj'')}_{J_2m_2(jj')J_3m_3(j'j'')}
\rightarrow 
{\cal{C}}^{J_1m_1q_1}_{J_2m_2q_2 J_3m_3q_3}
\label{limit of C hat}
\end{equation}
with the identification $j-j''=q_1$, $j-j'=q_2$ and $j'-j''=q_3$.
This relation can be proved by using the following asymptotic form
of the $6-j$ symbols. If $R \gg 1$, one obtains \cite{vmk}
\begin{equation}
\left\{
\begin{array}{ccc}
a   & b   & c \\
d+R & e+R & f+R \\
\end{array}
\right\}
\approx
\frac{(-1)^{a+b+c+2(d+e+f+R)}}{\sqrt{2R}}
\left(
\begin{array}{ccc}
a   & b   & c \\
e-f & f-d & d-e \\
\end{array}
\right),
\end{equation}
where the $3-j$ symbol is related to the Clebsch-Gordan 
coefficient as
\begin{equation}
\left(
\begin{array}{ccc}
J_1 & J_2 & J_3 \\
m_1 & m_2 & m_3 \\
\end{array}
\right)
=(-1)^{J_3+m_3+2J_1}\frac{1}{\sqrt{2J_3+1}}
C^{J_3m_3}_{J_1-m_1J_2-m_2}.
\end{equation}
The relation (\ref{limit of C hat}) implies 
that the fuzzy spherical harmonics $\hat{Y}_{Jm(jj')}$ 
give a matrix regularization of the monopole harmonics 
$\tilde{Y}_{Jmq}$ through the following correspondence:
\begin{eqnarray}
j-j' & \leftrightarrow & q, \nonumber\\
\hat{L}_A\circ & \leftrightarrow & L_A^{(q)}, \nonumber\\
\frac{1}{N_0}{\rm tr} & \leftrightarrow & \int\frac{d\Omega_2}{4\pi}.
\label{map between matrices and functions on S^2}
\end{eqnarray}
Furthermore, combining the above correspondence and the 
relations (\ref{integral of three Y tilde}) and  
(\ref{map between Y and Y tilde}),
we can also map the fuzzy spherical harmonics to the spherical 
harmonics on $S^3$.

\subsection*{A.4\hspace{0.5cm}Vector spherical harmonics}
We introduce vector spherical harmonics for three different types
of the spherical harmonics that we have defined above. 
The vector spherical harmonics are 
given by
\begin{align}
Y^{\rho}_{Jm\tilde{m}A}(\Omega_3)
&=i^{\rho}\sum_{n,p}U_{An}
C^{Qm}_{\tilde{Q}p\;1n}Y_{\tilde{Q}p\tilde{m}}(\Omega_3),
\nonumber\\
\tilde{Y}^{\rho}_{JmqA}(\Omega_2)
&=i^{\rho}\sum_{n,p}U_{An}
C^{Qm}_{\tilde{Q}p\;1n}\tilde{Y}_{\tilde{Q}pq}(\Omega_2),
\nonumber \\
\hat{Y}^{\rho}_{Jm(jj')A}
&=i^{\rho}\sum_{n,p}U_{An}
C^{Qm}_{\tilde{Q}p\;1n}\hat{Y}_{\tilde{Q}p(jj')},
\label{vector harmonics}
\end{align}
where $\rho = -1,0,1$ and
$Q=J+\delta_{\rho 1}$, $\tilde{Q}=J+\delta_{\rho \;-1}$.
These spherical harmonics transform as the vector representations
under $SU(2)$ rotation.
The unitary matrix $U$ is given by
\begin{equation}
U=\left(
\begin{array}{ccc}
-1 & 0 & 1 \\
-i & 0 & -i \\
0 & \sqrt{2} & 0 \\
\end{array}
\right). 
\end{equation}
The vector spherical harmonics satisfy
\begin{align}
&\frac{1}{\mu}\epsilon_{ABC}\nabla_{B}
Y^{\rho}_{Jm\tilde{m}C}=
i\epsilon_{ABC}{\cal L}_BY^{\rho}_{Jm\tilde{m}C}
+Y^{\rho}_{Jm\tilde{m}A}
=\rho(J+1)Y^{\rho}_{Jm\tilde{m}A},
\nonumber\\
&i\epsilon_{ABC}L_B^{(q)}\tilde{Y}^{\rho}_{JmqC}
+\tilde{Y}^{\rho}_{JmqA}
=\rho(J+1)\tilde{Y}^{\rho}_{JmqA},
\nonumber\\
&i\epsilon_{ABC}\hat{L}_B \circ \hat{Y}^{\rho}_{Jm(jj')C}
+\hat{Y}^{\rho}_{Jm(jj')A}
=\rho(J+1)\hat{Y}^{\rho}_{Jm(jj')A}.
\label{rotation}
\end{align}
The complex (hermitian) conjugates of these vector harmonics 
are evaluated as
\begin{align}
(Y^{\rho}_{Jm\tilde{m}A})^*
&=(-1)^{m-\tilde{m}+1}Y^{\rho}_{J-m-\tilde{m}A},
\nonumber\\
(\tilde{Y}^{\rho}_{JmqA})^*
&=(-1)^{m-q+1}\tilde{Y}^{\rho}_{J-m-qA},
\nonumber\\
(\hat{Y}^{\rho}_{Jm(jj')A})^{\dagger}
&=(-1)^{m-(j-j')+1}\hat{Y}^{\rho}_{J-m(j'j)A}.
\end{align}
The orthonormal relations are 
\begin{align}
\int \frac{d\Omega_3}{2\pi^2}
(Y^{\rho}_{Jm\tilde{m}A})^*Y^{\rho'}_{J'm'\tilde{m}'A}
&=
\delta_{JJ'}\delta_{mm'}
\delta_{\tilde{m}\tilde{m}'}\delta_{\rho\rho'},
\nonumber\\
\int \frac{d\Omega_2}{4\pi}
(\tilde{Y}^{\rho}_{JmqA})^*\tilde{Y}^{\rho'}_{J'm'qA}
&=
\delta_{JJ'}\delta_{mm'}\delta_{\rho\rho'},
\nonumber\\
\frac{1}{N_0} {\rm tr} \left(
(\hat{Y}^{\rho}_{Jm(jj')A})^{\dagger}\hat{Y}^{\rho'}_{J'm'(j'j)A}
\right)
&=
\delta_{JJ'}\delta_{mm'}\delta_{\rho\rho'}.
\end{align}
Finally, the integrals (or trace) of three vector harmonics
are given by
\begin{align}
\int \frac{d\Omega_3}{2\pi^2}
\epsilon_{ABC}
Y^{\rho_1}_{J_1m_1\tilde{m}_1A}
Y^{\rho_2}_{J_2m_2\tilde{m}_2B}
Y^{\rho_3}_{J_3m_3\tilde{m}_3C}
&={\cal E}_{
J_1m_1\tilde{m}_1\rho_1
J_2m_2\tilde{m}_2\rho_2
J_3m_3\tilde{m}_3\rho_3},
\nonumber\\
\int \frac{d\Omega_2}{4\pi}
\epsilon_{ABC}
\tilde{Y}^{\rho_1}_{J_1m_1q_1A}
\tilde{Y}^{\rho_2}_{J_2m_2q_2B}
\tilde{Y}^{\rho_3}_{J_3m_3q_3C}
&={\cal E}_{
J_1m_1q_1\rho_1
J_2m_2q_2\rho_2
J_3m_3q_3\rho_3},
\nonumber\\
\epsilon_{ABC}
\frac{1}{N_0} {\rm tr}
\left(
\hat{Y}^{\rho_1}_{J_1m_1(jj')A}
\hat{Y}^{\rho_2}_{J_2m_2(j'j'')B}
\hat{Y}^{\rho_3}_{J_3m_3(j''j)C}
\right)
&=\hat{{\cal E}}_{
J_1m_1(jj')\rho_1
J_2m_2(j'j'')\rho_2
J_3m_3(j''j)\rho_3},
\label{E}
\end{align} 
where the monopole charges must be conserved in 
the lefthand side of the second equality as $q_1+q_2+q_3=0$
and ${\cal E}$, $\hat{\cal E}$ are given by
\begin{align}
&{\cal E}_{
J_1m_1\tilde{m}_1\rho_1
J_2m_2\tilde{m}_2\rho_2
J_3m_3\tilde{m}_3\rho_3} \nonumber\\
&=
\sqrt{6(2J_1+1)(2J_1+2\rho_1^2+1)
       (2J_2+1)(2J_2+2\rho_2^2+1)
       (2J_3+1)(2J_3+2\rho_3^2+1)} \nonumber\\
&\;\;\;
\times (-1)^{-\frac{\rho_1+\rho_2+\rho_3+1}{2}}
\left\{
\begin{array}{ccc}
Q_1 & \tilde{Q}_1 & 1 \\
Q_2 & \tilde{Q}_2 & 1 \\
Q_3 & \tilde{Q}_3 & 1 
\end{array}
\right\}
\left(
\begin{array}{ccc}
Q_1 & Q_2 & Q_3 \\
m_1 & m_2 & m_3 
\end{array}
\right)
\left(
\begin{array}{ccc}
\tilde{Q}_1 & \tilde{Q}_2 & \tilde{Q}_3 \\
\tilde{m}_1 & \tilde{m}_2 & \tilde{m}_3 
\end{array}
\right),
\\
&\hat{{\cal E}}_{
J_1m_1(jj')\rho_1
J_2m_2(j'j'')\rho_2
J_3m_3(j''j)\rho_3} \nonumber\\
&=
\sqrt{6N_0(2J_1+1)(2J_1+2\rho_1^2+1)
       (2J_2+1)(2J_2+2\rho_2^2+1)
       (2J_3+1)(2J_3+2\rho_3^2+1)} \nonumber\\
&\;\;\;
\times (-1)^{-\frac{\rho_1+\rho_2+\rho_3+1}{2}
             -\tilde{Q}_1-\tilde{Q}_2-\tilde{Q}_3
             +2j+2j'+2j''}
\left\{
\begin{array}{ccc}
Q_1 & \tilde{Q}_1 & 1 \\
Q_2 & \tilde{Q}_2 & 1 \\
Q_3 & \tilde{Q}_3 & 1 
\end{array}
\right\}
\left(
\begin{array}{ccc}
Q_1 & Q_2 & Q_3 \\
m_1 & m_2 & m_3 
\end{array}
\right)
\left\{
\begin{array}{ccc}
\tilde{Q}_1 & \tilde{Q}_2 & \tilde{Q}_3 \\
j'' & j & j' 
\end{array}
\right\}.
\end{align}
As in (\ref{limit of C hat}), we can show 
\begin{equation}
\hat{{\cal E}}_{
J_1m_1(jj')\rho_1
J_2m_2(j'j'')\rho_2
J_3m_3(j''j)\rho_3}
\rightarrow
{\cal E}_{
J_1m_1\tilde{m}_1\rho_1
J_2m_2\tilde{m}_2\rho_2
J_3m_3\tilde{m}_3\rho_3},
\end{equation}
in the limit $N_0 \rightarrow \infty$ with 
$j-j'=q_1$, $j'-j''=q_2$ and $j''-j=q_3$ fixed.

\section{Derivation of (\ref{fields on base to those on total})}
\setcounter{equation}{0}
\renewcommand{\theequation}{B.\arabic{equation}}
In this appendix, we give the derivation of 
(\ref{fields on base to those on total}) in some detail.
\begin{align}
 [L_a,a_\alpha(x)]^{(s,t)}
 &=a_{\alpha,Jm}^{(q_{st})}(x)\otimes L_a\circ \hat{Y}_{Jm(j_sj_t)}
 \notag\\
 &=\int \frac{d\Omega_3}{2\pi^2} A_\alpha(z)
 Y_{Jmq_{st}}^\dagger (y)\otimes L_a\circ \hat{Y}_{Jm(j_sj_t)}
 \notag\\
 &=\int \frac{d\Omega_3}{2\pi^2} A_\alpha(z)
 \cL_a Y_{Jmq_{st}}^{\dagger}\otimes \hat{Y}_{Jm(j_sj_t)} 
 \notag\\
 &=\int \frac{d\Omega_3}{2\pi^2} \left(\cL_a A_\alpha(z)\right)
 Y_{Jmq_{st}}^\dagger \otimes \hat{Y}_{Jm(j_sj_t)}, \notag\\
\end{align}
where we have used (\ref{SU(2) alg. of S3 harmonics}) and (\ref{SU(2) algebra for fuzzy spherical harmonics}).
\begin{align}
 [\phi_a,\phi_b]^{(s,t)}
 &=\sum_{u}\left(\phi_{a,Jm}^{(q_{su})}\phi_{b,J'm'}^{(q_{ut})}
 -\phi_{b,Jm}^{(q_{su})}\phi_{a,J'm'}^{(q_{ut})}\right)
 \otimes \hat{Y}_{Jm(j_sj_u)}\hat{Y}_{J'm'(j_uj_t)} \notag\\
 &=\sum_{u}\left(\phi_{a,Jm}^{(q_{su})}\phi_{b,J'm'}^{(q_{ut})}
 -\phi_{b,Jm}^{(q_{su})}\phi_{a,J'm'}^{(q_{ut})}\right)
 \otimes \hat{\cC}^{J''m''(j_sj_t)}_{Jm(j_sj_u)\; J'm'(j_uj_t)}
 \hat{Y}_{J''m''(j_sj_t)} \notag\\
 &=
 \sum_{u,v}
 \int\frac{d\Omega_3}{2\pi^2}\frac{d\Omega_3'}{2\pi^2}
 \left\{ A_a(z) A_b(z') - A_b(z) A_a(z') \right\}
 Y_{Jmq_{su}}^\dagger(y)Y_{J'm'q_{vt}}^\dagger (y') \notag\\
 &\qquad\qquad \times
 \int \frac{d\Omega_3''}{2\pi^2}
 Y_{J''m''q_{st}}^\dagger(y'') Y_{Jmq_{su}}(y'')Y_{J'm'q_{vt}}(y'')
 \otimes \hat{Y}_{J''m''(j_sj_t)} \notag\\
 &=\int\frac{d\Omega_3}{2\pi^2}
 [A_a(z),A_b(z)]Y_{Jmq_{st}}^\dagger(y)\otimes \hat{Y}_{Jm(j_sj_t)}.
\end{align}
In the third and fourth lines of the righthand side,
we have used (\ref{limit of C hat}),
the charge conservation $\tilde{m}''=\tilde{m}+\tilde{m}'$ 
of $\cC^{J''m''\tilde{m}''}_{Jm\tilde{m}\; J'm'\tilde{m}'}$ and (\ref{integral of three Ys}), 
so that we have added the new summation over $v$ additionally
and replaced $q_{ut}$ by $q_{vt}$. Then, we can regard
the summation $\sum_{u,v}$ as $\sum_{q_{su},q_{vt}}$, and 
the last equality holds due to
(\ref{completeness condition of S3 harmonics}).

\section{Group manifold and coset space}
\setcounter{equation}{0}
\renewcommand{\theequation}{C.\arabic{equation}}
In this appendix, we describe some conventions 
on the group manifold $\tilde{G}$ and the coset space 
$\tilde{G}/H$ which we follow in this paper.

We parameterize an element of $\tilde{G}$ as
\begin{equation}
g(z)=L(x)h(y),
\label{g=Lh}
\end{equation}
where $L(x)\in \tilde{G}/H$, $h(y)\in H$,
the coordinates $z^M$, $x^{\mu}$ and $y^{m}$ 
parameterize $\tilde{G}$, $\tilde{G}/H$ and $H$
respectively and $z^M$ are decomposed into $(x^{\mu},y^i)$.
We can construct the right and left invariant 1-forms 
on $\tilde{G}$ as 
\begin{equation}
dgg^{-1}=-iE^A_RT^A, \;\;\; g^{-1}dg= iE_L^AT^A,
\label{left and right invariant 1-forms}
\end{equation}
where $A=1,\cdots ,{\rm dim}\tilde{G}$ and
$T^A$ represent the generators of $\tilde{G}$ which satisfy
the Lie algebra of $\tilde{G}$, $[T^A,T^B]=if_{ABC}T^C$. 
We decompose $T^A$ into $(T^{\alpha},T^a)$ where
$\alpha=1,\cdots,{\rm dim}\tilde{G}/H,\;
a={\rm dim}\tilde{G}/H+1,\cdots,{\rm dim}\tilde{G}$,
and we assume that $T^a$ satisfy the Lie algebra of $H$ which is
a subalgebra of $\tilde{G}$, $[T^a,T^b]=if_{abc}T^c$.
The both of $E^A_R$ and $E^A_L$ 
satisfy the Maurer-Cartan equation,
\begin{equation}
dE^A_R-\frac{1}{2}f_{ABC}E^B_R\wedge E^C_R=0,\;\;\;
dE^A_L-\frac{1}{2}f_{ABC}E^B_L\wedge E^C_L=0,
\label{Maurer-Cartan equation}
\end{equation}
We also introduce the right and left invariant 1-form for 
$L(x) \in \tilde{G}/H$ and $h(y)\in H$ as follows:
\begin{eqnarray}
dLL^{-1}=-i(e_R)^A_{\mu}(x)T^Adx^{\mu},\;\;\; 
L^{-1}dL= i(e_L)^A_{\mu}(x)T^Adx^{\mu},\nonumber\\
dhh^{-1}=-i(\tilde{e}_R)^a_m(y)T^ady^m, \;\;\; 
h^{-1}dh= i(\tilde{e}_L)^a_m(y)T^ady^m.
\end{eqnarray}
Then, we can write down the components of 
$E^A_R$ and $E_L^A$ explicitly: 
\begin{eqnarray}
(E_R)^A{}_M=\left(
\begin{array}{cc}
(e_R)^{\alpha}_{\mu} & {\rm Ad}(L)^{\alpha}{}_b(\tilde{e}_R)^b_m \\
(e_R)^a_{\mu} & {\rm Ad}(L)^a{}_b(\tilde{e}_R)^b_m
\end{array}
\right),  \;\;\;
(E_L)^A{}_M=\left(
\begin{array}{cc}
(e_L)^{\beta}_{\mu}{\rm Ad}(h)_{\beta \alpha} & 0 \\
(e_L)^{\beta}_{\mu}{\rm Ad}(h)_{\beta a} & (\tilde{e}_L)^a_m
\end{array}
\right), \label{components of E}
\end{eqnarray}
where ${\rm Ad}$ is defined as the adjoint action 
$gT^Ag^{-1}=T^B{\rm Ad}(g)_{BA}$.
The Cartan-Killing metric on $\tilde{G}$ is defined as 
\begin{equation}
ds^2=G_{MN}dz^Mdz^N=-2{\rm Tr}(dgg^{-1}dgg^{-1}).
\label{Cartan-Killing metric}
\end{equation}
In terms of the components (\ref{components of E}),
the above metric is written as
\begin{equation}
ds^2=(e_L)^{\alpha}_{\mu}(e_L)^{\alpha}_{\nu}dx^{\mu}dx^{\nu}
    +\{ (\tilde{e}_R)^a_mdy^m-(e_L)^a_{\mu}dx^{\mu} \}^2.
\label{metric on group manifold}
\end{equation}
We can regard the group manifold $\tilde{G}$ as the principal 
$H$ bundle on $\tilde{G}/H$.
By comparing (\ref{metric on group manifold}) and (\ref{metric}),
therefore, we can make the following identifications:
\begin{equation}
(e_L)^{\alpha}_{\mu}(e_L)^{\alpha}_{\nu}=g_{\mu\nu},\;
(e_L)^a_{\mu}=b_{\mu}^a,\; 
(\tilde{e}_R)^a_m(\tilde{e}_R)^a_n=h_{mn},
\end{equation}
where $g_{\mu\nu}$ and $h_{mn}$ are the metrics on $\tilde{G}/H$
and $H$, respectively, and $b_{\mu}^a$ are the local connection
1-forms of the principal $H$ bundle.
Namely, we can regard $(e_L)^{\alpha}_{\mu}$ and
$(\tilde{e}_R)^a_m$ as the vielbein on $\tilde{G}/H$ and 
$H$, respectively. 
The metric (\ref{Cartan-Killing metric}) is invariant 
under the right and left actions of $\tilde{G}$.
The corresponding right and left invariant Killing vectors on
$\tilde{G}$ are defined in terms of the inverse of $E^A$ as
\begin{equation}
{\cal L}^R_A=-i(E_R)_A^{M}\partial_{M}, \;\;\;
{\cal L}^L_A=-i(E_L)_A^{M}\partial_{M}.
\label{Killing vector on group manifold}
\end{equation}
By using (\ref{Maurer-Cartan equation}), we can show that
${\cal L}_A^R$ and ${\cal L}_A^L$ satisfy the Lie algebra of 
$\tilde{G}\times \tilde{G}$,
\begin{equation} 
[{\cal L}_A^R,{\cal L}_B^R]=if_{ABC}{\cal L}_C^R,\;\;\; 
[{\cal L}_A^L,{\cal L}_B^L]=if_{ABC}{\cal L}_C^L,\;\;\;
[{\cal L}_A^R,{\cal L}_B^L]=0,
\label{algebra of Killing vectors on group manifold}
\end{equation}
and they also satisfy the Killing vector equations,
\begin{equation}
\nabla_{M}{\cal L}_{AN}+\nabla_{N}{\cal L}_{AM}=0,
\end{equation}
where $\nabla_{M}$ are the covariant derivative on $\tilde{G}$ and
${\cal L}_{AM}=G_{MN}{\cal L}_{A}^N$.
We also define the following operators:
\begin{equation}
L_A=-i(E_R)_A^{\mu}\partial_{\mu}.
\label{Killing vector on coset space}
\end{equation}
One can show that $L_A$ do not depend on $y^m$ and 
they satisfy $[L_A,L_B]=if_{ABC}L_C$ by using 
(\ref{algebra of Killing vectors on group manifold}).
Furthermore, we can show that
\begin{equation}
\nabla^{(\tilde{G}/H)}_{\mu}L_{A\nu}+\nabla^{(\tilde{G}/H)}_{\nu}L_{A\mu}=0,
\end{equation}
where $L_{A\mu}=g_{\mu\nu} L_A^{\nu}$. 
Namely, $L_A$ are the Killing vectors on the coset space 
$\tilde{G}/H$.

\section{Metrics of $SU(3)$, $S^5$ and $CP^2$}
\setcounter{equation}{0}
\renewcommand{\theequation}{D.\arabic{equation}}
In this appendix, for concreteness, we give an explicit form of the metrics of $SU(3)$, $SU(3)/SU(2)\simeq S^5$
and $SU(3)/(SU(2)\times U(1)) \simeq CP^2$ \cite{Gerdt:2005xi}.
We parameterize an element $g$ of $SU(3)$ as 
\begin{align}
g=L(\chi,\theta,\varphi,\psi)Z(\tau)V(a,b,c),
\end{align}
where
\begin{align}
&L(\chi,\theta,\varphi,\psi)
=e^{i\varphi \lambda_3}e^{i\theta \lambda_2}e^{i\psi \lambda_3}e^{2i\chi\lambda_5}, \nonumber\\
&Z(\tau)=e^{-i\sqrt{3}(\tau-2\pi)\lambda_8}, \nonumber\\
&V(a,b,c)=e^{-ia\lambda_3}e^{-ib\lambda_2}e^{-ic\lambda_3},
\end{align}
and $0\leq\chi\leq\frac{\pi}{2}$, $0\leq\theta\leq\pi$, $0\leq\varphi < 2\pi$, $0\leq\psi < 4\pi$, 
$0\leq\tau < 2\pi$,
$0\leq a < 2\pi$, $0\leq b \leq \pi$ and $0\leq c < 4\pi$. $\lambda_1,\cdots,\lambda_8$ are the
Gell-Mann matrices and satisfy $\mbox{Tr}(\lambda_a\lambda_b)=\frac{1}{2}\delta_{ab}$. 
The metric of $SU(3)$ is given by
\begin{align}
dS_{SU(3)}^2 = &-\frac{1}{2}\mbox{Tr}(dgg^{-1}dgg^{-1}) \nonumber\\
             = &d\chi^2+\frac{1}{4}\sin^2\chi\left\{d\theta^2+\sin^2\theta d\varphi^2
                +\cos^2\chi(d\psi+\cos\theta d\varphi)^2\right\} \nonumber\\
             & +\frac{3}{4}\left\{d\tau+\frac{1}{2}\sin^2\chi(d\psi+\cos\theta d\varphi)\right\}^2 \nonumber\\
             &+\frac{1}{4}\left\{e^1+\cos\chi(\sin\psi d\theta-\sin\theta\cos\psi d\varphi)\right\}^2 \nonumber\\
             &+\frac{1}{4}\left\{e^2-\cos\chi(\cos\psi d\theta+\sin\theta\sin\psi d\varphi)\right\}^2 \nonumber\\
             &+\frac{1}{4}\left\{e^3-\frac{1}{2}(1+\cos^2\chi)(d\psi+\cos\theta d\varphi)\right\}^2,
\label{SU(3) metric}
\end{align}
where
\begin{align}
&e^1=-\sin a db+\cos a \sin b dc, \nonumber\\
&e^2=\cos a db+\sin a \sin b dc, \nonumber\\
&e^3=da +\cos b dc,
\end{align}
which are the right invariant 1-form of $SU(2)$.
$SU(3)$ is an $SU(2)\times U(1)$ bundle over $CP^2$.
The second line in the righthand side of (\ref{SU(3) metric}) is the Fubini-Study metric of $CP^2$.
The third line represents the $U(1)$ fiber structure while the fourth, fifth and sixth lines represent 
the $SU(2)$ fiber structure.
$SU(3)$ is also viewed as an $SU(2)$ bundle over $S^5\simeq SU(3)/SU(2)$.
The second and third lines together correspond to the metric of $S^5\simeq SU(3)/SU(2)$.
$S^5\simeq SU(3)/SU(2)$ is viewed as a $U(1)$ bundle over $CP^2$.
The metric of the ordinary unit $S^5$ is given by the sum of the second and third lines with the factor $3/4$ in
the third line replaced by $1/4$.

\section{Fuzzy $CP^n$}
\setcounter{equation}{0}
\renewcommand{\theequation}{E.\arabic{equation}}
In this appendix, we give a brief review of a construction
of fuzzy $CP^n$ 
\cite{CarowWatamura:2004ct,Alexanian:2001qj,Balachandran:2001dd,Kitazawa:2002xj,Grosse:2004wm,Dolan:2006tx}.

\subsection*{E.1\hspace{0.5cm}Functions on fuzzy $CP^n$}

Fuzzy $CP^n$ is a well-known example of noncommutative space
which is given by the quantization of coadjoint orbit of $SU(n+1)$
in terms of a certain matrix algebra acting on 
an appropriate representation space $V$. 
We can determine this matrix algebra and 
the representation space $V$ 
by matching the spectrum of 
functions on $CP^n$ and that on fuzzy $CP^n$.

In order to consider the spectrum of functions 
on $CP^n$,
We regard $CP^n$ as a coadjoint orbit in the Lie algebra of 
$SU(n+1)$.
\begin{equation}
CP^n = \{ \;gtg^{-1}\;|\;g \in SU(n+1) \} \simeq 
SU(n+1)/(SU(n)\times U(1)),
\label{coadjoint orbit}
\end{equation}
where $t$ is an element of the $SU(n+1)$ Lie algebra
such that
the stabilizer of $t$ is given by $SU(n)\times U(1)$.
For example, for the case of $CP^2$, we can 
take $t$ to be $\lambda_8$ which is invariant 
under $SU(2)\times U(1)$ adjoint action generated by
$\lambda_1,\lambda_2,\lambda_3$ and $\lambda_8$.
Functions on $CP^n$ should be invariant under the action of
$SU(n)\times U(1)$. 
Then, the space of functions on $CP^n$ is given by 
a direct sum of the representation spaces of $SU(n+1)$
which contain $SU(n) \times U(1)$ invariant states:
\begin{equation}
C^{\infty}(CP^n)=\bigoplus_{J=0}^{\infty} V_{[J,0,\cdots, 0,J]}
\label{KK modes on CP^n}
\end{equation}
where we denote 
$[J,0,\cdots, 0,J]$ as the Dynkin index of $SU(n+1)$,
and $V_{[J,0,\cdots, 0,J]}$ represents the corresponding 
irreducible representation space of the $SU(n+1)$ Lie algebra.
One can show that $V_{[J,0,\cdots, 0,J]}$ are 
the only spaces which contain the $SU(n) \times U(1)$ singlets. 

The space of functions on fuzzy $CP^n$ is obtained by 
introducing a cutoff $\Lambda$ in (\ref{KK modes on CP^n}) as 
\begin{equation}
\bigoplus_{J=0}^{\Lambda} V_{[J,0,\cdots, 0,J]}
=V_{[\Lambda,0,\cdots,0]} \otimes V_{[\Lambda,0,\cdots,0]}^*.
\label{spectrum on fuzzy CP^n}
\end{equation}
By definition, it is obvious that the above spectrum 
on fuzzy $CP^n$ tends to the spectrum 
(\ref{KK modes on CP^n}) on $CP^n$
in the commutative limit $\Lambda \rightarrow \infty$.
Note that the righthand side of the 
above equation can be viewed as a space of matrices. 
From this viewpoint, we make an identification 
$V=V_{[\Lambda,0,\cdots,0]}$ and 
regard functions on fuzzy $CP^n$ as matrices acting 
on the vector space $V$.
In particular, 
the coordinates on fuzzy $CP^n$ are identified with 
\begin{equation}
\hat{\xi}_A= \hat{L}_A^{[\Lambda,0,\cdots,0]},
\label{coordinates on fuzzy CP^n}
\end{equation}
which are the generators of $SU(n+1)$ in 
the irreducible representation specified by the
Dynkin index $[\Lambda,0,\cdots,0]$.  
These coordinates on fuzzy $CP^n$ are actually reduced to the 
coordinates on $CP^n$ in the commutative limit 
through a map which will be defined 
in the last part of this section.

\subsection*{E.2\hspace{0.5cm}Derivatives on fuzzy $CP^n$}

In order to construct differential operators on fuzzy $CP^n$,
let us recall the simplest case of fuzzy $CP^1 \simeq S^2$.
In this case, we established the differential operators on 
fuzzy $S^2$ in appendix A. 
As shown in
(\ref{map between matrices and functions on S^2}),
the adjoint action 
of the $SU(2)$ generators is reduced to
the action of the Killing vectors on $S^2$ in the 
commutative limit. 
We can generalize this fact into the case of 
fuzzy $CP^n$ with $n\geq 2$. 
The adjoint action of the $SU(n+1)$ generators on the 
space of square matrices (\ref{spectrum on fuzzy CP^n}), 
$[\hat{L}_A^{[\Lambda,0,\cdots,0]},\;\cdot\;]$,
is mapped into the action of the Killing vectors 
on the space of functions on $CP^n$ in the commutative limit.

\subsection*{E.3\hspace{0.5cm}$U(1)$ monopoles on fuzzy $CP^n$}
 
Topologically nontrivial field configurations including
$U(1)$ monopoles can be realized on fuzzy $CP^n$.
If we consider rectangular matrices in addition to the square 
matrices (\ref{spectrum on fuzzy CP^n}),
the concept of fiber bundles naturally arises.
Let us again consider the case of fuzzy $CP^1$.
We have shown in appendix A 
that the basis of $(2j+1)\times (2j'+1)$ 
rectangular matrices, $\hat{Y}_{Jm(jj')}$, are mapped into
local sections of the $U(1)$ fiber bundle on $S^2$. 
In this correspondence,
The difference $j-j'$ is identified with the monopole charge 
$q$ of the $U(1)$ bundle.
This fact is also generalized into the case of 
$CP^n$ with $n\geq 2$.
For the case of $CP^n$,
we consider a space of rectangular matrices, 
\begin{equation}
V_{[\Lambda+q,0,\cdots,0,]} \otimes V_{[\Lambda-q,0,\cdots,0,]}^*.
\label{rectangular matrices}
\end{equation}
Here, the charge $q$ is a half integer and 
we take $\Lambda\pm q$ to be integers.
When $q=0$, $\Lambda$ is an integer and this is the case of square
matrices (\ref{spectrum on fuzzy CP^n}).
We can show that 
elements of (\ref{rectangular matrices}) are mapped into 
local sections of $U(1)$ fiber bundle on $CP^n$ with
the monopole charge $q$. 
Furthermore, we can extend the action of 
the differential operators 
$[\hat{L}_A^{[\Lambda,0,\cdots,0]},\;\cdot\;]$ discussed above
to the action on rectangular matrices as follows.
\begin{equation}
\hat{L}_A \circ \hat{M}_q =
\hat{L}_A^{[\Lambda+q,0,\cdots,0]} \hat{M}_q
-\hat{M}_q\hat{L}_A^{[\Lambda-q,0,\cdots,0]},
\label{derivatives for rectangular matrices}
\end{equation}
where $\hat{M}_q$ is an element of (\ref{rectangular matrices}).
When $q=0$, $\hat{M}_0$ is just a square matrix and 
$\hat{L}_A \circ$ are nothing but the commutators 
$[\hat{L}_A^{[\Lambda,0,\cdots,0]},\;\cdot\;]$.
The operators $\hat{L}_A \circ$ map the space 
(\ref{rectangular matrices}) to itself and 
they are reduced to the angular momentum operators
in the presence of a $U(1)$ monopole with the magnetic 
charge $q$ in the commutative limit.
We will show these facts in the following subsections.

\subsection*{E.4\hspace{0.5cm}Fock space representation}

In order to construct a map between matrices and functions 
on $CP^n$, we introduce the Fock space representation 
developed in \cite{Dolan:2006tx}.
Let $a^{\dagger}_{\alpha}$, $\alpha=1,2,\cdots,n+1$ be a set of
creation operators and $a^{\alpha}$ be a set of 
annihilation operators
which annihilate the Fock vacuum $|0\rangle$. They satisfy
the Heisenberg commutation relations.
\begin{equation}
[a^{\alpha},a^{\beta}]=[a^{\dagger}_{\alpha},a^{\dagger}_{\beta}]=0,
\;\;\; 
[a^{\alpha},a^{\dagger}_{\beta}]=\delta_{\beta}^{\alpha}.
\label{Heisenberg commutation relations}
\end{equation}
By acting the creation operators on the vacuum state $|0\rangle$,
we can construct the entire Fock space ${\cal F}$ which is spanned by
\begin{equation}
  | p_1, p_2, \cdots ,p_{n+1} \rangle
 =\frac{1}{\sqrt{p_1!p_2!\cdots p_{n+1}!}}
 (a_1^{\dagger})^{p_1}(a_2^{\dagger})^{p_2}
 \cdots (a_{n+1}^{\dagger})^{p_{n+1}} 
 |0 \rangle.
\end{equation} 
In terms of the operators
(\ref{Heisenberg commutation relations}), 
we can construct elements 
of the Lie algebra of $SU(n+1)$,
\begin{equation}
\hat{L}_A= a^{\dagger}_{\alpha} 
(T_A)^{\alpha}{}_{\beta} a^{\beta},
\label{generators in Fock space}
\end{equation}
where $T_A=\hat{L}_A^{[1,0,\cdots,0]}$ represent 
the generators of $SU(n+1)$ in the fundamental 
representation.
We also define the number operator which commutes
with all the operators in (\ref{generators in Fock space}).
\begin{equation}
\hat{N}=a^{\dagger}_{\alpha}a^{\alpha}.
\label{number operator}
\end{equation}
The operators (\ref{generators in Fock space}) and 
(\ref{number operator}) act on the Fock space ${\cal F}$,
and satisfy
\begin{equation}
[\hat{L}_A,\hat{L}_B]=if_{ABC}\hat{L}_C,
\;\;\;[\hat{L}_A, \hat{N}]=0.
\end{equation}
We can decompose the Fock space ${\cal F}$ into 
the eigenspaces of $\hat{N}$ as 
\begin{equation}
{\cal F}=\bigoplus_{p=0}^\infty V_{[p,0,\cdots,0]},
\end{equation}
where $p$ represent an eigenvalue of $\hat{N}$. 
The basis of each eigenspace $V_{[p,0,\cdots,0]}$ is formed by
\begin{equation}
| {\bm  \alpha } _p \rangle =
| \alpha_1, \alpha_2, \cdots ,\alpha_p \rangle
 =\frac{1}{\sqrt{p!}}
 a_{\alpha_1}^{\dagger}a_{\alpha_2}^{\dagger}
 \cdots a_{\alpha_p}^{\dagger} 
 |0 \rangle,
\end{equation}
where ${\bm \alpha}_p$ is an abbreviation of a set of $p$ indices,
$(\alpha_1, \alpha_2, \cdots ,\alpha_p)$.

Let us consider square matrices which are elements of 
(\ref{spectrum on fuzzy CP^n}).
These matrices are generally written as 
\begin{equation}
\hat{M}=\hat{M}^{{\bm \alpha}_{\Lambda}}
  {}_{{\bm\beta}_{\Lambda}}
  | {\bm \alpha}_{\Lambda} \rangle
  \langle {\bm\beta}_{\Lambda} |.
\end{equation}
We define a new basis of these matrices 
to see the correspondence with the spectrum of functions on 
$CP^n$, (\ref{KK modes on CP^n}):
\begin{equation}
\hat{Y}_{{\bm \beta}_J}{}^{{\bm \alpha}_J}
=N_{\Lambda J}^n {\cal P}
_{{\bm \beta}_J,{\bm \tau}_J}
{}^{{\bm \alpha}_J,{\bm \sigma}_J}
| {\bm \sigma}_J,{\bm \gamma}_{\Lambda-J} \rangle
\langle {\bm \tau}_J,{\bm \gamma}_{\Lambda-J}|,
\label{fuzzy harmonics on CP^n}
\end{equation}
where $\Lambda-J$ indices ${\bm \gamma}_{\Lambda-J}$
are contracted and
${\cal P}_{{\bm \beta}_J,{\bm \tau}_J}
{}^{{\bm \alpha}_J,{\bm \sigma}_J}$ 
is the projection operator 
onto the representation space $V_{[J,0,\cdots,0,J]}$ which appeared
in the decomposition (\ref{spectrum on fuzzy CP^n}),
that is, 
it removes all traces between ${\bm \alpha}_J$ and ${\bm \beta}_J$.
For example, 
\begin{align}
\hat{Y}= N_{\Lambda 0}^n {\bm 1} , \;\;\;
\hat{Y}_{\beta}{}^{\alpha} = N_{\Lambda 1}^n \left(
|\beta, {\bm \gamma}_{\Lambda -1}\rangle
\langle \alpha, {\bm \gamma}_{\Lambda-1}|
-\frac{1}{2}\delta_{\beta}^{\alpha}{\bm 1}
\right).
\end{align}
Hence, $\hat{Y}_{{\bm \beta}_J}{}^{{\bm \alpha}_J}$ belong 
to the representation $V_{[J,0,\cdots,0,J]}$ and they are
mapped to the corresponding spherical harmonics on $CP^n$ 
in the commutative limit which are elements of 
(\ref{KK modes on CP^n}).
$N_{\Lambda J}^n$ is an appropriate normalization constant
which is determined by the following orthonormality of
the basis,
\begin{equation}
{\rm tr}\left( 
(\hat{Y}_{{\bm \beta}_J}{}^{{\bm \alpha}_J})^{\dagger}
\hat{Y}_{{\bm \tau}_{J'}}{}^{{\bm \sigma}_{J'}}
\right)
=\delta_{JJ'}{\cal P}
_{{\bm \alpha}_J,{\bm \tau}_J}
{}^{{\bm \beta}_J,{\bm \sigma}_J}.
\end{equation}
In the case of $n=1$, 
$\hat{Y}_{{\bm \beta}_J}{}^{{\bm \alpha}_J}$ are essentially
the same as the fuzzy spherical harmonics which are defined in
(\ref{fuzzy spherical harmonics}).
The action of differential operators on fuzzy $CP^n$
is given by the adjoint action of operators in 
(\ref{generators in Fock space}). Then, one can 
evaluate the eigenvalues of the Laplacian as follows:
\begin{equation}
[\hat{L}_A,[\hat{L}_A,
\hat{Y}_{{\bm \beta}_J}{}^{{\bm \alpha}_J}]]
=
J(J+n)\hat{Y}_{{\bm \beta}_J}{}^{{\bm \alpha}_J}.
\end{equation}   
The above spectrum completely matches the spectrum of 
functions on $CP^n$ up to the cutoff $\Lambda$.

In terms of the Fock space representation, 
we can also express rectangular matrices which are 
elements of (\ref{rectangular matrices}).
Those rectangular matrices are generally expressed as
\begin{equation}
\hat{M}_q=(\hat{M}_q)^{{\bm \alpha}_{\Lambda+q}}
  {}_{{\bm\beta}_{\Lambda-q}}
  | {\bm \alpha}_{\Lambda+q} \rangle
  \langle {\bm\beta}_{\Lambda-q} |.
\label{Fock space representation for M_q}
\end{equation}
These matrices are expanded by a similar basis to 
(\ref{fuzzy harmonics on CP^n}).
Note that the direct product representation 
$(\ref{rectangular matrices})$ is decomposed as
\begin{equation}
\bigoplus_{J=|q|}^{\Lambda}V_{[J+q,0,\cdots,0,J-q]}.
\label{sections on fuzzy CP^n}
\end{equation}
For each representation space 
in (\ref{sections on fuzzy CP^n}) with fixed $J$,
we can use the following basis:
\begin{equation}
\hat{Y}^{(q)}{}_{{\bm \beta}_{J+q}}{}^{{\bm \alpha}_{J-q}}
=N_{\Lambda Jq}^n {\cal P}
_{{\bm \beta}_{J+q},{\bm \tau}_{J-q}}
{}^{{\bm \alpha}_{J-q},{\bm \sigma}_{J+q}}
| {\bm \sigma}_{J+q},{\bm \gamma}_{\Lambda-J} \rangle
\langle {\bm \tau}_{J-q},{\bm \gamma}_{\Lambda-J}|.
\label{fuzzy monopole harmonics on CP^n}
\end{equation}
As in the case of square matrices,  
${\cal P}_{{\bm \beta}_{J+q},{\bm \tau}_{J-q}}
{}^{{\bm \alpha}_{J-q},{\bm \sigma}_{J+q}}$
is a projection operator onto the space 
(\ref{sections on fuzzy CP^n}) with fixed $J$
and $N_{\Lambda Jq}^n$ is a normalization constant 
which is determined by 
\begin{equation}
{\rm tr}\left( 
(\hat{Y}^{(q)}{}_{{\bm \beta}_{J+q}}{}^{{\bm \alpha}_{J-q}})^{\dagger}
\hat{Y}^{(q)}{}_{{\bm \tau}_{J'+q}}{}^{{\bm \sigma}_{J'-q}}\right)
=\delta_{JJ'}
{\cal P}_{{\bm \tau}_{J+q},{\bm \alpha}_{J-q}}
{}^{{\bm \sigma}_{J-q},{\bm \beta}_{J+q}}.
\end{equation}
When $q=0$, 
$\hat{Y}^{(0)}{}_{{\bm \beta}_J}{}^{{\bm \alpha}_J}$
are identical with the square matrices 
(\ref{fuzzy harmonics on CP^n}).
The action of differential operators on 
$ \hat{Y}^{(q)}{}_{{\bm \beta}_{J+q}}{}^{{\bm \alpha}_{J-q}}$
is given by (\ref{derivatives for rectangular matrices}).
We can evaluate the eigenvalues of the Laplacian as follows:
\begin{equation}
(\hat{L}_A\circ)^2 
\hat{Y}^{(q)}{}_{{\bm \beta}_{J+q}}{}^{{\bm \alpha}_{J-q}}
=\left(J(J+1)+\frac{n-1}{n+1}q^2\right)
\hat{Y}^{(q)}{}_{{\bm \beta}_{J+q}}{}^{{\bm \alpha}_{J-q}}.
\end{equation}
The above spectrum is the same as the spectrum of local sections
of $U(1)$ bundle on $CP^n$ up to the cutoff.
We show in the following that the rectangular matrices 
$ \hat{Y}^{(q)}{}_{{\bm \beta}_{J+q}}{}^{{\bm \alpha}_{J-q}}$
are indeed mapped to the local sections on $CP^n$.

\subsection*{E.5\hspace{0.5cm}Relation between matrices and sections}
Let us recall the spherical harmonics on $CP^n$. 
In a spinorial basis, they are given by 
\begin{equation}
\tilde{Y}_{{\bm \beta}_J}{}^{{\bm \alpha}_J}
=N_J^n {\cal P}
_{{\bm \beta}_J,{\bm \tau}_J}
{}^{{\bm \alpha}_J,{\bm \sigma}_J}
\bar{w}_{\sigma_1}\cdots\bar{w}_{\sigma_J}
w^{\tau_1}\cdots w^{\tau_J},
\label{spherical harmonics on CP^n}
\end{equation}
In the above expression,
$w^{\alpha}$ are the coordinates of 
$S^{2n+1}\simeq SU(n+1)/SU(n)$ 
which satisfy $\sum_{\alpha} |w^{\alpha}|^2=1$ and
the normalization constant $N_J^n$ is determined by 
\begin{equation}
\int_{CP^n} \omega^n \; 
(\tilde{Y}_{{\bm \beta}_J}{}^{{\bm \alpha}_J})^*
\tilde{Y}_{{\bm \tau}_{J'}}{}^{{\bm \sigma}_{J'}}
=\delta_{JJ'}{\cal P}
_{{\bm \alpha}_J,{\bm \tau}_J}
{}^{{\bm \beta}_J,{\bm \sigma}_J},
\end{equation}
where $\omega^n$ is the volume form on $CP^n$.
The functions (\ref{spherical harmonics on CP^n}) are
invariant under the $U(1)$ phase rotation so that they 
can be regarded as global sections (functions) on $CP^n$.
We can generalize (\ref{spherical harmonics on CP^n})
to a basis of local sections of the $U(1)$ 
monopole bundle on $CP^n$. 
The local sections of the monopole bundle with the 
magnetic charge $q$ can be expanded by 
\begin{equation}
\tilde{Y}^{(q)}{}_{{\bm \beta}_{J+q}}{}^{{\bm \alpha}_{J-q}}
=N_{Jq}^n {\cal P}
_{{\bm \beta}_{J+q},{\bm \tau}_{J-q}}
{}^{{\bm \alpha}_{J-q},{\bm \sigma}_{J+q}}
\bar{w}_{\sigma_1}\cdots\bar{w}_{\sigma_{J+q}}
w^{\tau_1}\cdots w^{\tau_{J-q}},
\label{monopole harmonics on CP^n}
\end{equation}
which are normalized as
\begin{equation}
\int_{CP^n} \omega^n \; 
(\tilde{Y}^{(q)}{}_{{\bm \beta}_{J+q}}{}^{{\bm \alpha}_{J-q}})^*
\tilde{Y}^{(q)}{}_{{\bm \tau}_{J'+q}}{}^{{\bm \sigma}_{J'-q}}
=\delta_{JJ'}{\cal P}
_{{\bm \alpha}_{J-q},{\bm \tau}_{J+q}}
{}^{{\bm \beta}_{J+q},{\bm \sigma}_{J-q}}.
\end{equation}
$\tilde{Y}^{(q)}{}_{{\bm \beta}_{J+q}}{}^{{\bm \alpha}_{J-q}}$
are not invariant 
under the $U(1)$ phase rotation, so that they transform 
as the local sections of the monopole bundle on $CP^n$ with the 
magnetic charge $q$. Note that
$\tilde{Y}^{(0)}{}_{{\bm \beta}_J}{}^{{\bm \alpha}_J}$
are nothing but the global sections,
$\tilde{Y}_{{\bm \beta}_J}{}^{{\bm \alpha}_J}$.

The relation between matrices and sections on $CP^n$ 
is given by the diagonal coherent state map \cite{Dolan:2006tx}. 
Let us consider a matrix $\hat{M}_q$ which is an element of 
(\ref{rectangular matrices}) and expanded as in
(\ref{Fock space representation for M_q}).
$\hat{M}_q$ corresponds to a section 
of the monopole bundle on $CP^n$ through the map.
In particular, when $q=0$, $\hat{M}_0$ is just a square matrix 
and corresponds to a global section on $CP^n$.
The map to the sections is given by
\begin{equation}
\tilde{M}_q(w,\bar{w})=
\langle w, \Lambda+q | \hat{M}_q | w,\Lambda-q \rangle,
\label{map between matrices and functions}
\end{equation}
where
\begin{equation}
| w, p \rangle = \frac{1}{\sqrt{p !}}
(w^{\alpha}a_{\alpha}^{\dagger})^p |0 \rangle .
\end{equation}
The map (\ref{map between matrices and functions}) 
is equivalent to the following replacement up to
an over all constant factor,
\begin{eqnarray}
(a^{\dagger}_{\alpha})^L \rightarrow \bar{w}_{\alpha},\;\;\;\;
(a^{\alpha})^L \rightarrow \frac{\partial}{\partial \bar{w}_{\alpha}},
\nonumber\\
(a^{\alpha})^R \rightarrow w^{\alpha},\;\;\;\;
(a^{\dagger}_{\alpha})^R \rightarrow \frac{\partial}{\partial w^{\alpha}},
\end{eqnarray}
where the superscripts $L$ and $R$ express that 
the operators act on matrices from the left and right, respectively.
Through this correspondence,
(\ref{fuzzy harmonics on CP^n}) and 
(\ref{fuzzy monopole harmonics on CP^n}) are
mapped to (\ref{spherical harmonics on CP^n}) and
(\ref{monopole harmonics on CP^n}) respectively.
Furthermore, the differential operators $\hat{L}_{A}\circ$ 
on fuzzy $CP^n$ are 
mapped to
\begin{equation}
\hat{L}_{A}\circ \rightarrow 
L_A^{(q)}=
\bar{w}_{\alpha}(T_A)^{\alpha}{}_{\beta}
\frac{\partial}{\partial \bar{w}_{\beta}}
-w^{\alpha}(T_A^*)_{\alpha}{}^{\beta}
\frac{\partial}{\partial w^{\beta}}.
\label{Angular momentum operators on $CP^n$}
\end{equation}
When $q=0$, 
these operators
act on the functions (\ref{spherical harmonics on CP^n}) and
they can be identified with the Killing vectors on $CP^n$.
In the case $q \neq 0$, however, 
they act on the local sections 
(\ref{monopole harmonics on CP^n}) so that 
the derivative along the $U(1)$ fiber direction does 
not vanish and yields additional terms which are
proportional to the charge $q$.
In this case, the operators 
(\ref{Angular momentum operators on $CP^n$}) 
can be interpreted as the angular momentum operators on $CP^n$ in
the presence of a monopole with the charge $q$.

\end{document}